\documentclass[12pt]{article}
\usepackage{amsmath,amsfonts,amssymb}
\usepackage{graphicx}
\usepackage{enumerate}
\usepackage{natbib}
\usepackage{url} 

\usepackage{bm,bbm}
\usepackage[dvipsnames]{xcolor}
\usepackage{subfig,booktabs}
\usepackage{paralist}
\usepackage{color}
\allowdisplaybreaks
\numberwithin{equation}{section}

\renewcommand{\Pr}{\mathbb{P}}
\newcommand{\reals}{\mathbb{R}} 
\newcommand{\E}{\mathbb{E}}
\newcommand{\diff}{\, \mathrm{d}}

\newcommand{\I}{\mathbbm{1}}

\newcommand{\DC}{D\setminus C}

\newcommand{\GP}{\operatorname{GP}}

\newcommand{\bzero}{\bm{0}}
\newcommand{\bone}{\bm{1}}

\newcommand{\bgamma}{{\bm{\gamma}}}

\newcommand{\bEta}{{\bm{\eta}}}

\newcommand{\bsigma}{{\bm{\sigma}}}

\newcommand{\bu}{\bm{u}}

\newcommand{\bw}{\bm{w}}
\newcommand{\bx}{\bm{x}}

\newcommand{\bR}{\bm{R}}
\newcommand{\bV}{\bm{V}}

\newcommand{\bX}{\bm{X}}
\newcommand{\bY}{\bm{Y}}

\newcommand{\blind}{0}

\begin{document}

\def\spacingset#1{\renewcommand{\baselinestretch}%
{#1}\small\normalsize} \spacingset{1}


\if0\blind
{
 \title{\bf Peaks over thresholds modelling with multivariate generalized Pareto distributions}
  \author{ 
   Anna Kiriliouk \hspace{.2cm} \\
    \small Erasmus University Rotterdam \\
    \small Erasmus School of Economics \\
    \small 3000 DR Rotterdam, the Netherlands.\\
    \small E-mail: kiriliouk@ese.eur.nl \\
   \and
    Holger Rootz\'{e}n \hspace{.2cm}\\
     \small Chalmers University of Technology \\
    \small  Department of Mathematical Sciences \\
    \small SE-412 96 Gothenburg, Sweden.\\
    \small E-mail: hrootzen@chalmers.se \\
    \and
        Johan Segers \hspace{.2cm}\\
    \small Universit\'{e} catholique de Louvain\\
    \small Institut de Statistique, Biostatistique \\ 
    \small et Sciences Actuarielles\\
    \small Voie du Roman Pays~20 \\
    \small 1348 Louvain-la-Neuve, Belgium.\\
    \small E-mail: johan.segers@uclouvain.be\\
    \and
   Jennifer L. Wadsworth \hspace{.2cm}\\
     \small Lancaster University \\
    \small  Department of Mathematics and Statistics \\
    \small Fylde College LA1 4YF, Lancaster, England. \\
    \small E-mail: j.wadsworth@lancaster.ac.uk \\
    }
    \date{}
\maketitle
} \fi

\if1\blind
{
  \bigskip
  \bigskip
  \bigskip
  \begin{center}
    {\LARGE\bf Peaks over thresholds modelling with multivariate generalized Pareto distributions}
\end{center}
  \medskip
} \fi

\bigskip
\begin{abstract}

 
When assessing the impact of extreme events, it is often not just a single component, but the combined behaviour of several components which is important. Statistical modelling using multivariate generalized Pareto (GP) distributions constitutes the multivariate analogue of univariate peaks over thresholds modelling, which is widely used in finance and engineering. We develop general methods for construction of multivariate GP distributions and use them to create a variety of new statistical models. A censored likelihood procedure is proposed to make inference on these models, together with a threshold selection procedure, goodness-of-fit diagnostics, and a computationally tractable strategy for model selection. The models are fitted to returns of stock prices of four UK-based banks and to rainfall data in the context of landslide risk estimation. Supplementary materials and codes are available online.
\end{abstract}

\noindent%
\vfill
\hfill {\tiny technometrics tex template (do not remove)}

\newpage
\spacingset{1.45} 

\section{Introduction}
\label{sec:intro}

Univariate peaks over thresholds modelling with the generalized Pareto (GP) distribution is extensively used in hydrology to quantify risks of extreme floods, rainfalls and waves  \citep{katz+parlange+naveau:2002,hawkes+gouldy+tawn+owen:2002}. It is the standard way to estimate Value at Risk in financial engineering \citep{McNeil+Frey+Embrechts:2015}, and has been useful in a wide range of other areas, including  wind engineering, loads on structures, strength of materials, and traffic safety \citep{Ragan+Manuel:2008, anderson+demare+rootzen:2013, Gordonetal:2013}.

However often it is the flooding of not just one but  many dikes which determines the damage caused by a big flood, and a flood in turn may be caused by rainfall in not just one but in several catchments. Financial risks typically are not determined by the behaviour of one financial instrument, but by many instruments which together form a financial portfolio. Similarly, in the  other areas listed above it is often multivariate rather than univariate modeling which is required.

There is a growing body of probabilistic literature devoted to multivariate GP distributions \citep{rootzen2006,falk2008,ferreira2014,rootzen2016,rootzen2017}. To our knowledge, however, there are only a few papers that use these as a statistical model \citep{thibaud2015,huser2015,defondeville2017}, and these only use a single family of GP distributions. 

In this paper we advance the practical usefulness of multivariate peaks over threshold modelling by developing  general construction methods of multivariate GP distributions and by using them  to create a variety of new GP distributions. To facilitate practical use, we suggest computationally tractable strategies for model selection, demonstrate model fitting via censored likelihood, and provide techniques for threshold selection and model validation. 

 We illustrate the new methods by using them to derive multivariate risk estimates for returns of stock prices of four UK-based banks (Section~\ref{sec:banks}), and show that these can be  more useful for  portfolio risk management than currently available one-dimensional estimates.  Environmental risks often involve physical constraints not taken into account by available methods. We  estimate landslide risks using models which handle such constraints, thereby providing more realistic estimates (Section \ref{sec:rainfall}).

The new parametric multivariate GP models are given in Sections~\ref{sec:models} and \ref{sec:examples}, and the model selection, fitting, and validation methods are developed in Section~\ref{sec:inference}. An important feature is that we can estimate marginal and dependence parameters simultaneously, so that confidence intervals include the full estimation uncertainty. We also give some background needed for the use of the models (Section \ref{sec:background}).

The ``point process method''  \citep{coles1991} provides an alternative approach for modelling threshold exceedances. However, the multivariate GP distribution has practical and conceptual advantages, in so much as it is a proper multivariate distribution. It also separates modelling of the times  of threshold exceedances  and the distribution of the threshold excesses in a useful way.


We limit ourselves to the situation where all components show full asymptotic dependence. Technically, with this we mean that the margins of the multivariate GP distribution do not put any mass on their lower endpoints. The contrary case, which requires detecting subgroups of variables which show full asymptotic dependence, constitutes a challenging area for future research, especially when the number of variables is large.

The inference method that we propose is based on likelihoods for data points that are censored from below, so as to avoid bias resulting from inclusion of observations that are not high enough to warrant the use of the multivariate GP distribution. The formulas of the censored likelihoods for the parametric models that we propose are given in the online supplementary material. In that supplement, which includes all \textsf{R} codes, we also report on bivariate tail dependence coefficients, further numerical experiments illustrating the models and the model choice procedure, and we give further details on the case studies.

\section{Background}
\label{sec:background}

This section provides a brief overview of basic properties of multivariate GP distributions, as needed for understanding and practical use. Let $\bm{Y}$ be a random vector in $\mathbb{R}^d$ with distribution function $F$. A common assumption on $\bm{Y}$ is that it is in the so-called \emph{max-domain of attraction} of a multivariate max-stable distribution, $G$. This means that if $\bY_1,\ldots,\bY_n$ are independent and identically distributed copies of $\bY$, then one can find sequences $\bm{a}_n \in(0,\infty)^d$ and $\bm{b}_n \in \mathbb{R}^d$ such that
\begin{align}
 \Pr[\{\max_{1\leq i\leq n}\bm{Y}_i - \bm{b}_n\}/\bm{a}_n \leq \bm{x}] \to G(\bm{x}), \label{eq:maxstabconv}
\end{align}
with $G$ having non-degenerate margins.
In~\eqref{eq:maxstabconv} and throughout, operations involving vectors are to be interpreted componentwise. If convergence~\eqref{eq:maxstabconv} holds, then
\begin{align}
\max\left\{\frac{\bm{Y}-\bm{b}_n}{\bm{a}_n}, \bEta\right\} \mid \bm{Y} \not\leq \bm{b}_n \overset{d}{\to} \bm{X}, \qquad \text{as } n \rightarrow \infty, \label{eq:gpdconv}
\end{align}
where $\bX$ follows a multivariate GP distribution \citep{rootzen2016}, and where $\bEta$ is the vector of lower endpoints of the GP distribution, to be given below. We let $H$ denote the distribution function of $\bm{X}$, and $H_1,\ldots,H_d$ its marginal distributions. Typically the margins $H_j$ are not univariate GP, due to the difference between the conditioning events $\{Y_j>b_{n,j}\}$ and $\{\bm{Y}\not\leq \bm{b}_n\}$ in the one-dimensional and $d$-dimensional limits. Still, the marginal distributions conditioned to be positive are GP distributions. That is, writing $a_+ = \max(a, 0)$, we have
\begin{align}\label{eq:margins}
\overline{H}_j^+ (x) := \Pr[X_j > x \mid X_j > 0] = (1+\gamma_j x /\sigma_j)^{-1/\gamma_j}_+,
\end{align}
where $\sigma_j$ and $\gamma_j$ are marginal scale and shape parameters. The unconditional margins $H_j$ have lower endpoints $\eta_j=-\sigma_j/\gamma_j$ if $\gamma_j>0$ and $\eta_j=-\infty$ otherwise. The link between $H$ and $G$ is $H(\bm{x}) =  \{\log G(\min(\bx,\bzero))-\log G(\bx)\}/\{\log G(\bzero)\}$, and we say that $H$ and $G$ are \emph{associated}.

Following common practice in the statistical modelling of extremes, $H$ may be used as a model for data which arise as multivariate excesses of high thresholds. Hence, if $\bm{u}\in \mathbb{R}^d$ is a threshold vector that is ``sufficiently high'' in each margin, then we approximate $\bm{Y} - \bm{u} \mid \bm{Y} \not\leq \bm{u}$  by a member $\bX$ of the class of multivariate GP distributions,  with $\bm{\sigma}$, $\bm{\gamma}$, the marginal exceedance probabilities $\Pr(Y_j > u_j)$, and the dependence structure to be estimated. In practice the truncation by the vector $\bEta$ in \eqref{eq:gpdconv} is only relevant when dealing with mass on lower-dimensional subspaces, and is
 outside the scope of the present paper. Observe that there is no difficulty in directly considering large values of $\bY$ itself, i.e., the conditional distribution of $\bY$ given that $\bY \nleq \bu$, by changing the support to $\{\bm{x}:\bm{x}\nleq \bm{u}\}$; this is equivalent to replacing $\bx$ by $\bx - \bu$ in density~\eqref{eq:mgpdens} below.
 
 By straightforward computation, the distribution function of componentwise maxima of a Poisson number of GP variables for $\bx \geq \bzero$ equals $\exp\{-t(1-H(\bx))\}$, which is the max-stable distribution $G^t$, and where $t$ is the mean of the Poisson distribution. Hence, a peaks over thresholds analysis, combined with estimation of the occurrence rate of events, also provides an estimate of the joint distribution of, say, yearly maxima.
 
 The following are further useful properties of GP distributions; for details and proofs we refer to \citet{rootzen2016} and \citet{rootzen2017}.

\noindent
{\bf  Threshold stability:}
 GP distributions are \emph{threshold stable}, meaning that if $\bX \sim H$ follows a GP distribution and if $\bw\geq \bzero$, with $H(\bw)<1$ and $\bsigma+\bgamma\bw > \bzero$, then
 \begin{align*}
  \bX-\bw \mid \bX \not\leq \bw
  \text{ is GP with parameters } \bm{\sigma} + \bm{\gamma} \bw \text{ and } \bm{\gamma}.
 \end{align*}
Hence if the thresholds are increased, then the distribution of conditional excesses is still GP, with a new set of scale parameters, but retaining the same vector of shape parameters. The practical relevance of this stability is that the model form does not change at higher
levels, which is useful for extrapolating further into the tail.

A special role is played by the levels $\bw = \bw_t := \bsigma(t^\bgamma -1)/\bgamma$: these have the stability property that for any set
$A \subset \{\bx \in \mathbb{R}^d: \bx \nleq \bzero \}$
it holds that, for $t \geq 1$,
 \begin{equation}\label{eq:stability}
\Pr[\bX \in \bw_t + t^\bgamma A] = \Pr[\bX \in A] / t,
 \end{equation}
 where $\bw_t + t^\bgamma A = \{\bw_t + t^\bgamma  \bx: \bx \in A\}$. This follows from equation~\eqref{eq:gpconstr} along with the representation of $\bm{X}_0$ to be given in equation~\eqref{eq:X0constr}. The $j$-th component of $\bw_t$, $\sigma_j(t^{\gamma_j} -1)/\gamma_j$, is the $1-1/t$ quantile of $H_j^+$. Equation~\eqref{eq:stability} provides one possible tool for checking if a multivariate GP distribution is appropriate; see Section~\ref{sec:diagnostics}.

\noindent
{\bf Lower dimensional conditional margins:}
 Lower dimensional margins of GP distributions are typically not GP. Instead $\bm{X}_{J} \mid \bm{X}_{J} \not\leq \bzero_J$ does follow a GP distribution, for $\bm{X}_J = (x_j : j\in J)$ and $J \subset\{1,\ldots, d\}$. Combined with the threshold stability property above, we also have that if $\bw_J \in\mathbb{R}^{|J|}$ is such that $\bw_J \geq \bzero$, $H_J(\bw_J)<1$ and $\bsigma_J + \bgamma_J \bw_J > \bzero$ then $\bX_J -\bw_J  \mid \bX_J \not\leq \bw_J $ follows a GP distribution.


\noindent
{\bf Sum-stability under shape constraints:}
If $\bX$ follows a multivariate GP distribution, with scale parameter $\bm{\sigma}$ and shape parameter $\bm{\gamma} = \gamma \bm{1}$, then for weights $a_j>0$ such that $\sum_{j=1}^d a_j X_j > 0$ with positive probability, we have
\begin{align} 
 \textstyle\sum_{j=1}^d a_j X_j \mid  \sum_{j=1}^d a_j X_j >0 \sim \GP(\sum_{j=1}^d a_j \sigma_j, \gamma). \label{eq:sumstab}
\end{align}
Thus weighted sums of components of a multivariate GP distribution with equal shape parameters, conditioned to be positive, follow a univariate GP distribution with the same shape parameter and with  scale parameter equal to the weighted sum of the marginal scale parameters. This in particular may be useful for financial modelling. Equation~\eqref{eq:sumstab} holds regardless of the particular GP dependence structure.
However, the probability of the conditioning event, $\{\sum_{j=1}^{d} a_jX_j > 0\}$, will differ for different dependence structures.

\section{Model construction}
\label{sec:models}

We use three constructions to develop general parametric classes of GP densities, labelled $h_{\bm{T}}, h_{\bm{U}},$ and $h_{\bm{R}}$. For the first two, one first constructs a standard form density for a variable $\bX_0$ with $\bsigma =\bone,\bm{\gamma} = \bm{0}$, and then obtains a density on the observed scale through the standard  transformation
\begin{align}
\bX \overset{\mathrm{d}}{=} \bm{\sigma}\frac{e^{\bm{\gamma} \bX_0}-\bm{1}}{\bm{\gamma}} \label{eq:gpconstr},
\end{align}
with the distribution $\bX$ supported on $\{\bm{x} \in \mathbb{R}^d: \bm{x}\not\leq \bm{0} \}$. For $\gamma_j = 0$, the corresponding component of the right-hand side of equation~\eqref{eq:gpconstr} is simply $\sigma_j X_{0,j}$. The third class of densities, $h_{\bm{R}}$, is constructed directly on the observed scale. Each of the constructions starts with choosing a suitable probability distribution, $\bm{T}$, $\bm{U}$,  or $\bm{R}$, the ``generator'' of the class, which is combined with a common random intensity, or strength, to yield the GP model. More details, alternative constructions, and intuition for the three forms are given in \cite{rootzen2016,rootzen2017}.

We note that several articles have previously used random vectors to generate dependence structures for extremes, e.g. \citet{segers2012}, \cite{thibaud2015} and \citet{aulbach2015}, whilst the literature on max-stable modelling for spatial extremes also relies heavily on this device \citep{dehaan1984, schlather2002, davison2012}. However, it is only recently that these constructions have led to simple density formulas for GP distributions \citep{rootzen2017}, which we exploit to build several new models. Explicit forms for a number of useful GP densities are given in Section~\ref{sec:examples}; here we discuss their construction further.



\noindent
{\bf Standard form densities.}

We first focus on how to construct suitable densities for the random vector $\bX_0$, which, through equation~\eqref{eq:gpconstr}, lead to densities for the multivariate GP distribution with marginal parameters $\bsigma$ and $\bgamma$. Let $E$ be a unit exponential random variable and let $\bm{T}$ be a $d$-dimensional random vector, independent of $E$. Define $\max(\bm{T}) = \max_{1 \le j \le d} T_j$. Then the random vector
\begin{align}
 \bX_0 = E + \bm{T} - \max(\bm{T}) \label{eq:X0constr}
\end{align}
is a GP vector with support included in the set $\{ \bm{x} \in \reals^d : \bm{x} \nleq \bzero \}$ and with $\bsigma = \bm{1}$ and $\bgamma = \bzero$ (interpreted as the limit for $\gamma_j \to 0$ for all $j$). Moreover, \emph{every} such GP vector can be expressed in this way \citep{ferreira2014,rootzen2016}. The probability of the $j$-th component being positive is $\Pr[ X_{0,j} > 0 ] = \E[ e^{T_j - \max(\bm{T})} ]$, which, in terms of the original data vector $\bm{Y}$, corresponds to the  probability $\Pr[ Y_j > u_j \mid \bm{Y} \nleq \bm{u} ]$, i.e., the probability that the $j$-th component exceeds its corresponding threshold given that one of the $d$ components does.

Suppose $\bm{T}$ has a density $f_{\bm{T}}$ on $(-\infty,\infty)^d$. By Theorem~5.1 of \citet{rootzen2016}, the density of $\bX_0$ is given by
\begin{equation}
\label{eq:casef}
   h_{\bm{T}}(\bx;\bm{1},\bm{0})
   = \frac{\I\{\max(\bm{x})>0\}}{e^{\max(\bm{x})}} \int_{0}^\infty f_{\bm{T}} (\bx+ \log t) \, t^{-1} \diff t.
\end{equation}
One way to construct models therefore is to assume distributions for $\bm{T}$ which provide flexible forms for $h_{\bm{T}}$, and for which ideally the integral in~\eqref{eq:casef} can be evaluated analytically.

One further construction of GP random vectors is given in \citet{rootzen2016}. If $\bm{U}$ is a $d$-dimensional random vector with density $f_{\bm{U}}$ and such that $\E[e^{U_j}] < \infty$ for all $j = 1, \ldots, d$, then the following function also defines the density of a GP distribution:
\begin{align}
  \label{eq:caseg}
  h_{\bm{U}} (\bx;\bm{1},\bm{0})
  = \frac{\I\{\max(\bm{x})>0\}}{\E[e^{\max(\bm{U})}]}
  \int_{0}^\infty  f_{\bm{U}} (\bx+ \log t)  \diff t.
\end{align}
The marginal exceedance probabilities are now $\Pr[ X_{0,j} > 0 ] = \E[ e^{U_j} ] / \E[ e^{\max(\bm{U})} ]$. Formulas \eqref{eq:casef} and \eqref{eq:caseg} can be obtained from one another via a change of measure.

Where $f_{\bm{T}}$ and $f_{\bm{U}}$ take the same form, then the similarity in integrals between~\eqref{eq:casef} and~\eqref{eq:caseg} means that if one can be evaluated, then typically so can the other; several instances of this are given in the models presented in Section~\ref{sec:examples}. What is sometimes more challenging is calculation of the normalization constant $\E[e^{\max(\bm{U})}] = \int_0^\infty \Pr[ \max(\bm{U}) > \log t ] \diff t$ in~\eqref{eq:caseg}. Nonetheless, the model in~\eqref{eq:caseg} has the particular advantage over that of~\eqref{eq:casef} that it behaves better across various dimensions: if the density of the GP vector $\bm{X}$ is $h_{\bm{U}}$ and if $J \subset \{1, \ldots, d\}$, then the density of the GP subvector $\bm{X}_J \mid \bm{X}_J \nleq \bzero_J$ is simply $h_{\bm{U}_J}$. This property is advantageous when moving to the spatial setting, since the model retains the same form when numbers of sites change, which is useful for spatial prediction.

\noindent
{\bf Densities after transformation to the observed scale.}

The densities above are in the standardized form $\bm{\sigma}=\bm{1}$, $\bm{\gamma}=\bm{0}$.
Using \eqref{eq:gpconstr}, we obtain  general  densities which are approximations to the conditional density of $\bY-\bu$ given that $\bY \nleq \bu$, for  the original data $\bY$:
\begin{align}
 h(\bx;\bm{\sigma}, \bm{\gamma}) =
 h\left(\tfrac{1}{\bgamma}\log(1+\bm{\gamma}\bx /\bm{\sigma}) ; \bm{1},\bm{0}\right)
 \prod_{j=1}^d \frac{1}{\sigma_j + \gamma_j x_j}.
 \label{eq:mgpdens}
\end{align}
In~\eqref{eq:mgpdens}, $h$ may be either $h_{\bm{T}}$ or $h_{\bm{U}}$. 

\noindent
{\bf Densities constructed on observed scale.}

The models \eqref{eq:mgpdens} are built on a standardized scale, and then transformed to the observed, or ``real'' scale. Alternatively, models can be constructed directly on the real scale, which gives the possibility of respecting structures, say additive structures, in a way which is not possible with the other two models; this approach will be used to model ordered data in Section~\ref{sec:rainfall}. One way of presenting this is to define the random vector $\bm{R}$ in terms of $\bm{U}$ in~\eqref{eq:caseg} through the componentwise transformation
\begin{equation}
  \label{eq:ru}
  R_j =
  \begin{cases}
    (\sigma_j/\gamma_j) \exp (\gamma_j U_j), & \gamma_j \neq 0, \\
    \sigma_jU_j, & \gamma_j = 0,
  \end{cases}
\end{equation}
and develop suitable models for $\bR$. This gives the GP density
\begin{equation}\label{eq:realdens}
h_{\bm{R}} (\bx ; \bm{\sigma}, \bm{\gamma} ) =
\frac{\I \left\{ \max(\bm{x}) > 0 \right\}}{\E [e^{\max(\bm{U})}]}
  \int_0^\infty
    t^{\sum_{j=1}^d \gamma_j}
    f_{\bm{R}} \left( \bigl( g(t;x_j,\sigma_j,\gamma_j) \bigr)_{j=1}^d \right)
  \diff t,
\end{equation}
where $f_{\bm{R}}$ denotes the density of $\bm{R}$ and where
\begin{align*}
  g(t;x_j,\sigma_j,\gamma_j) =
  \begin{cases}
    t^{\gamma_j} \left(x_j + \sigma_j/\gamma_j \right), & \gamma_j \neq 0, \\
    x_j+ \sigma_j \log t, & \gamma_j = 0.
  \end{cases}
\end{align*}
The $d$ components of $\bm{U}$ are found by inverting equation~\eqref{eq:ru}. For $\bm{\sigma} = \bm{1}$ and $\bm{\gamma} = \bm{0}$, the densities \eqref{eq:caseg} and \eqref{eq:realdens} are the same.

In light of the abundance of possibilities, we note the following, which may help the user to select a suitable model: Computation, and particularly simulation, is simplest for the $h_{\bm{T}}$ densities, and these models are continuous at $\gamma_j =0$, for each $j$. However, spatial prediction and lower dimensional margins are unnatural for this model class. Instead,  prediction, spatial modelling, and lower dimensional margins work well for the $h_{\bm{U}}$ densities, and this model class is also continuous  at $\gamma_j =0$. Finally, for the $h_{\bm{R}}$ class, prediction, spatial modelling, and lower dimensional margins are also natural, and the class additionally permits more physically realistic modelling. However, it is not continuous at $\gamma_j =0$.

\section{Likelihood-based inference}
\label{sec:inference}

Working within a likelihood-based framework for inference allows many benefits. Firstly, comparison of nested models can be done using likelihood ratio tests. This is important as the number of parameters can quickly grow large if margins and dependence are fitted simultaneously, allowing us to test for simplifications in a principled manner. Secondly, incorporation of covariate effects is straightforward in principle. For univariate peaks over thresholds, such ideas were introduced by \citet{davison1990}, but nonstationarity in dependence structure estimation has received comparatively little attention. Thirdly, such likelihoods could also be exploited for a Bayesian approach to inference if desired.

\subsection{Censored likelihood}
\label{sec:censored}

The density~\eqref{eq:mgpdens} is the basic ingredient in a likelihood. However, we will use~\eqref{eq:mgpdens} as a contribution only when all components of the observed translated vector $\bm{Y}-\bm{u}$ are ``large'', in the sense of exceeding a threshold $\bm{v}$, with $\bm{v}\leq \bm{0}$. Where some components of $\bm{Y}-\bm{u}$ fall below $\bm{v}$, the contribution is censored in those components. The reasoning for this is twofold:
\begin{enumerate}
 \item For $\gamma_j>0$, the lower endpoint of the multivariate GP distribution is $-\sigma_j/\gamma_j$. Censored likelihood avoids small values of a component affecting the fit too strongly.
 \item Without censoring, bias in the estimation of parameters controlling the dependence can be larger than that for censored estimation, see \citet{huser2015}.
\end{enumerate}
Censored likelihood for inference on extreme value models was first used by \cite{smith1997} and \cite{ledford1997}, and is now a standard approach to enable more robust inference.
Let  $C\subset D=\{1,\ldots,d\}$ contain the indices for which components of $\bm{Y}-\bm{u}$ fall below the corresponding component of $\bm{v}$, i.e., $Y_j -u_j\leq v_j$ for $j\in C$, and $Y_j-u_j>v_j$ for $j\in D\setminus C$, with at least one such $Y_j>u_j$. For each realization of $\bm{Y}$, we use the likelihood contribution
\begin{align}
\label{eq:hC}
  h^C(\bm{y}_{\DC}-\bm{u}_{\DC},\bm{v}_C;\bm{\sigma},\bm{\gamma})
  = \int_{\times_{j\in C}(-\infty,u_j+v_j]}  h(\bm{y}-\bm{u};\bm{\sigma}, \bm{\gamma}) \diff \bm{y}_{C},
\end{align}
with $\bm{y}_{C}=(y_j)_{j \in C}$, which is equal to~\eqref{eq:mgpdens} with $\bm{x}=\bm{y}-\bm{u}$ if $C$ is empty, i.e., if all components $y_j>u_j+v_j$. The supplementary material contains forms of censored likelihood contributions for the models presented in Section~\ref{sec:examples}. For $n$ independent observations $\bm{y}_1,\ldots,\bm{y}_n$ of $\bm{Y} \mid \bm{Y}\not\leq\bm{u}$, the  censored likelihood function to be optimized is
\begin{align}
\label{eq:lhood}
 L(\bm{\theta},\bm{\sigma}, \bm{\gamma}) = \prod_{i=1}^n h^{C_i}(\bm{y}_{i, D \setminus C_i}-\bm{u}_{D \setminus C_i},\bm{v}_{C_i};\bm{\theta}, \bm{\sigma},\bm{\gamma}),
\end{align}
where $C_i$ denotes the censoring subset for $\bm{y}_i$, which may be empty, and $\bm{\theta}$ represents parameters related to the model that we assumed for the generator.

\subsection{Model choice}
\label{sec:modelchoice}

When fitting multivariate GP distributions to data on the observed scale we have a large variety of potential models and parameterizations. For non-nested models, Akaike's Information Criterion (AIC = $-2~\times $ log-likelihood + $2~\times $ number of parameters) can be used to select a model with a good balance between parsimony and goodness-of-fit. When looking at nested models, e.g., to test for simplifications in parameterization, we can use likelihood ratio tests. Because of the many possibilities for model fitting, we propose the following model-fitting strategy to reduce the computational burden, which we will employ in Section~\ref{sec:banks}.
\begin{compactenum}[(i)]
 \item Standardize the data to common exponential margins, $\bm{Y}_E$, using the rank transformation (i.e., the probability integral transform using the empirical distribution function);
 \item select a multivariate threshold, denoted $\bm{u}$ on the scale of the observations, and $\bm{u}_E$ on the exponential scale, using the method of Section~\ref{sec:diagnostics};
 \item fit the most complicated standard form model within each class (i.e., maximum number of possible parameters) to the standardized data $\bm{Y}_E-\bm{u}_E \mid \bm{Y}_E\not\leq\bm{u}_E$;
 \item select as the standard form model class the one which produces the best fit to the standardized data, in the sense of smallest AIC;
 \item use likelihood ratio tests to test for simplification of models within the selected standard form class, and select a final standard form model; \label{test}
 \item fit the GP margins simultaneously with this standard form model, to $\bm{Y}-\bm{u} \mid \bm{Y}\not\leq\bm{u}$ by maximizing~\eqref{eq:lhood};
 \item Use likelihood ratio tests to find  simplifications in the marginal parameterization.
\end{compactenum}
Although this strategy is not guaranteed to result in a final GP model that is globally optimal, in the sense of minimizing an information criterion such as AIC, it should still result in a sensible model whilst avoiding enumeration and fitting of an unfeasibly large number of possibilities. The goodness of  fit of the final model can be checked via diagnostic plots and tests (hereafter ``diagnostics'').

\subsection{Threshold selection and model diagnostics}
\label{sec:diagnostics}
An important issue that pervades extreme value statistics --- in all dimensions --- is the selection of a threshold above which the limit model provides an adequate approximation of the distribution of threshold exceedances. Here this amounts to ``how can we select a vector $\bm{u}$ such that $\bm{Y} - \bm{u} \mid\bm{Y}\not\leq\bm{u}$ is well-approximated by a GP distribution?''. There are two considerations to take into account: $Y_j - u_j \mid Y_j>u_j$ should be well-approximated by a univariate GP distribution, for $j=1,\ldots,d$, and the dependence structure of $\bm{Y} - \bm{u} \mid\bm{Y}\not\leq\bm{u}$ should be well-approximated by that of a multivariate GP distribution.
Marginal threshold selection has a large body of literature devoted to it; see \citet{scarrott2012} and \citet{caeiro2016} for recent reviews. Threshold selection for dependence models is a much less well studied problem. Contributions include \citet{lee2015} who considers threshold selection via Bayesian measures of surprise, and \citet{wadsworth2016}  who examines how to make better use of so-called parameter stability plots, offering a method that can be employed on any parameter, pertaining to the margins or dependence structure. Recently, \citet{wan2017} proposed a method based on asessing independence between radial and angular distributions.

Here we propose exploiting the stability property of multivariate GP distributions, and use the measure of asymptotic dependence
\begin{equation*}\label{eq:chi1d}
\chi_{1:d}(q):=\frac{\Pr[F_1(Y_1)> q, \ldots, F_d(Y_d)> q]}{1-q}, 
\end{equation*}
where $Y_j\sim F_j$ and the related quantity for the limiting GP distribution
\begin{equation*}
\chi_{H}(q) :=\frac{\Pr[H_1(X_1)>q,\ldots,H_d(X_d)>q]}{1-q},~~~q\in(0,1) \label{eq:chiq}
\end{equation*}
to guide threshold selection for the dependence structure. For a suitable choice of $A$, property~\eqref{eq:stability} implies that $\chi_{H}(q)$ is constant for sufficiently large $q$ such that $H_j (X_j) > q$ implies $X_j > 0$ for $j \in \{1,\ldots,d\}$.

If $\bm{Y} \sim F$ and  $\bm{Y}-\bm{u} \mid \bm{Y}\nleq\bm{u} \sim H$, then on the region $q>\max_j F_j(u_j)$, we have $\chi_{1:d}(q) = \chi_H(q')$ with $q'=\{q-F(\bm{u})\}/\{1-F(\bm{u})\}$. A consequence of this is that $\chi_{1:d}(q)$ should be constant on the region $\bm{Y}>\bm{u}$, if $\bm{u}$ represents a sufficiently high dependence threshold. The empirical version  $\widehat{\chi}_{1:d} (q)$  of $\chi_{1:d}(q)$ is defined by
\begin{equation}\label{eq:chiemp}
\widehat{\chi}_{1:d} (q) := \frac{\sum_{i=1}^n \mathbbm{1} \left\{ \widehat{F}_1 (Y_1) >q, \ldots, \widehat{F}_d (Y_d) > q \right\}}{n (1-q)}, \qquad q \in [0,1),
\end{equation}
where $\widehat{F}_1,\ldots,\widehat{F}_d$ represent the empirical distribution functions. If we use \eqref{eq:chiemp}
to identify $q^*= \inf\{0 < \tilde{q} < 1 : \chi_{1:d} (q) \equiv \chi~\forall~q>\tilde{q}\}$, then $\bm{u}=(F_{1}^{-1}(q^*),\ldots, F_{d}^{-1}(q^*))$ should provide an adequate threshold for the dependence structure. Once suitable thresholds have been identified for margins, $\bm{u}_{\mathrm{m}}$, and dependence, $\bm{u}_{\mathrm{d}}$, then a threshold vector which is suitable for the entire multivariate model is $\bm{u}=\max(\bm{u}_{\mathrm{m}},\bm{u}_{\mathrm{d}})$.

Having identified a multivariate GP model and a threshold above which to fit it, a key concern is to establish whether the goodness-of-fit is adequate.
For the dependence structure, one diagnostic comes from comparing $\widehat{\chi}_{1:d} (q)$ for $q \rightarrow 1$ to its theoretical limit $\chi_{1:d}$,
which for models $h_{\bm{T}}$ in \eqref{eq:casef} has the form
$
\chi_{1:d} = \E\left[\min_{1\leq j \leq d} \{e^{T_j - \max(\bm{T})}/\E(e^{T_j - \max(\bm{T})})\}\right],
$
whilst for models $h_{\bm{U}}$ in \eqref{eq:caseg} we get
$
 \chi_{1:d} = \E\left[\min_{1\leq j \leq d} \{e^{U_j}/\E(e^{U_j})\}\right].
$
The form of $\chi_{1:d}$ for $h_{\bm{R}}$ models  follows through equation~\eqref{eq:ru}.
In some cases these expressions may be obtained analytically, but they can always be evaluated  by simulation \citep{rootzen2016}.

A further diagnostic uses that $\Pr[X_j>0 ] = \E[e^{T_j-\max(\bm{T})}] = \E[e^{U_j}]/\E[e^{\max( \bm{U} )}]$. Thus, one compares $\Pr[Y_j > u_j] / \Pr[\bm{Y}\not\leq\bm{u}]$ with the relevant model-based probability. These are the same for each margin when the $u_j$ are equal marginal quantiles.

Equation~\eqref{eq:stability} suggests a model-free diagnostic of whether a multivariate GP model may be appropriate. To exploit this, one defines a set of interest $A$, and compares the number of points of $\bm{Y}-\bm{u} \mid \bm{Y}\not\leq\bm{u}$ that lie in $A$ to $t$ times the number of points of $(\bm{Y}-\bm{u}-\bm{w}_t)/t^{\bm{\gamma}} \mid \bm{Y}\not\leq\bm{u}$ lying in $A$ for various choices of $t > 1$. According to~\eqref{eq:stability}, the ratio of these numbers should be approximately equal to $1$. Note that setting $A=\{\bx : \bx>\bzero\}$ is equivalent to computing $\chi_H$ with $H_1,\ldots,H_d$ replaced by $H^+_1,\ldots,H^+_d$.

Finally, in the event that the margins can be modelled with identical shape parameters, one can test property~\eqref{eq:sumstab} by examining the adequacy of the implied univariate GP distribution from a multivariate fit.

\section{UK bank returns}
\label{sec:banks}

We examine  weekly negative raw returns on the prices of the stocks from four large UK banks: HSBC (H), Lloyds (L),  RBS (R) and Barclays (B). Data were downloaded from Yahoo Finance. Letting $Z_{j,t}$, $j\in\{H,L,R,B\}$, denote the closing stock price (adjusted for stock splits and dividends) in week $t$ for bank $j$, the data we examine are the negative returns $Y_{j,t} = 1-Z_{j,t}/Z_{j,t-1}$, so that large positive values of $Y_{j,t}$ correspond to large relative losses for that stock. The observation period is 10/29/2007 -- 10/17/2016, with $n=470$ datapoints. The data are unfiltered, i.e., heteroscedasticity has not been removed. This is because we are not trying to predict at specific time points, but rather understand the global extremal dependence.

\begin{figure}[ht]
\centering
\includegraphics[width=0.24\textwidth]{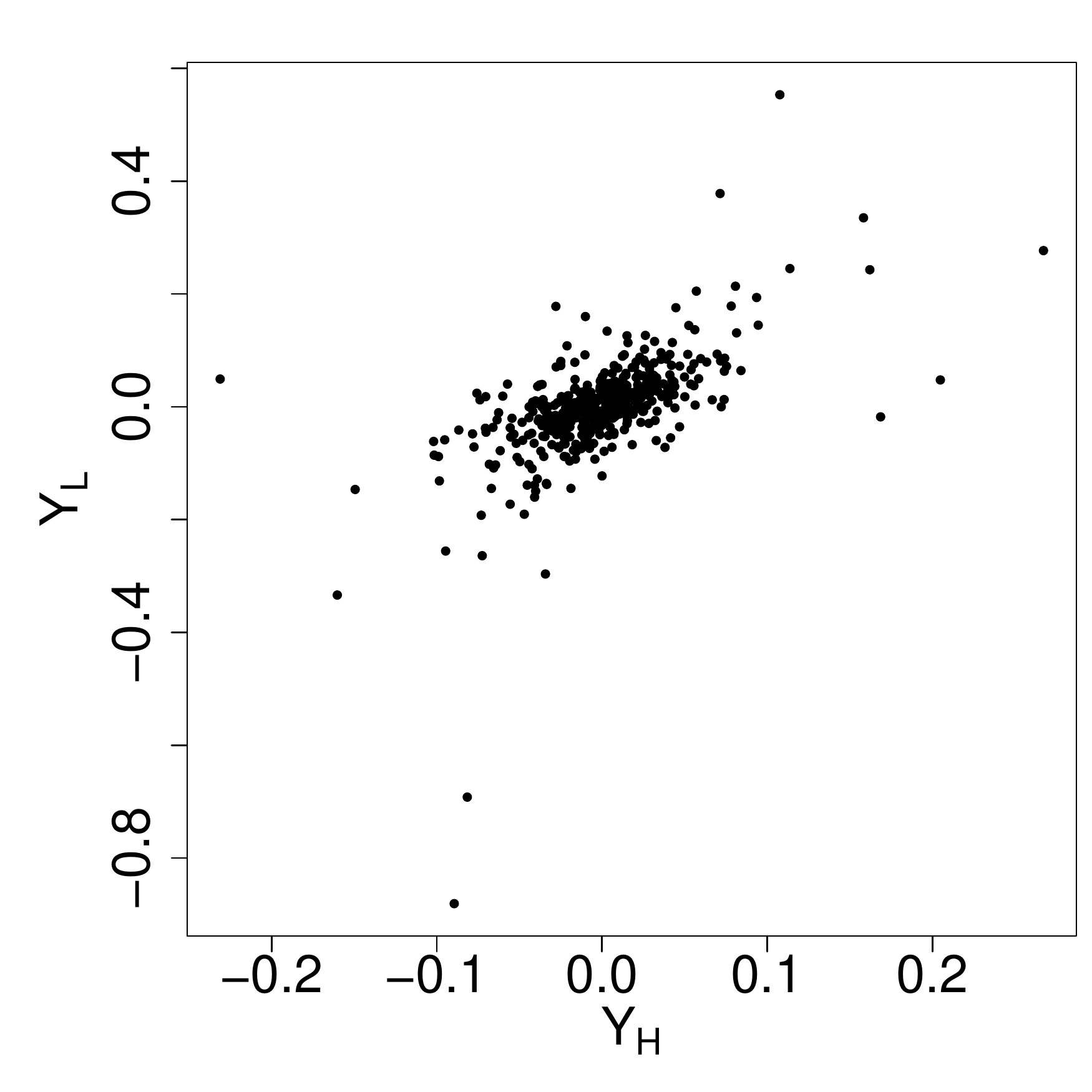}
\includegraphics[width=0.24\textwidth]{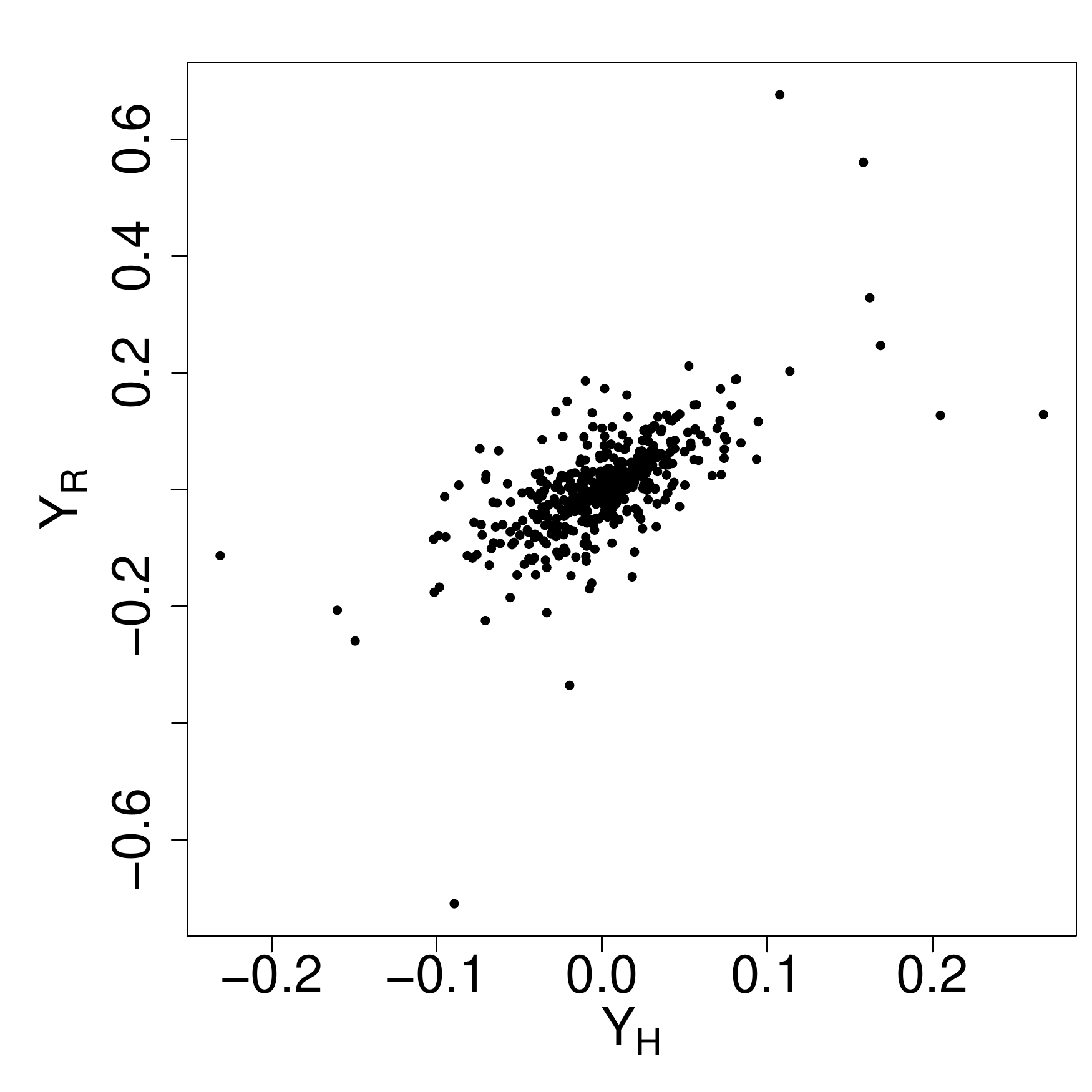}
\includegraphics[width=0.24\textwidth]{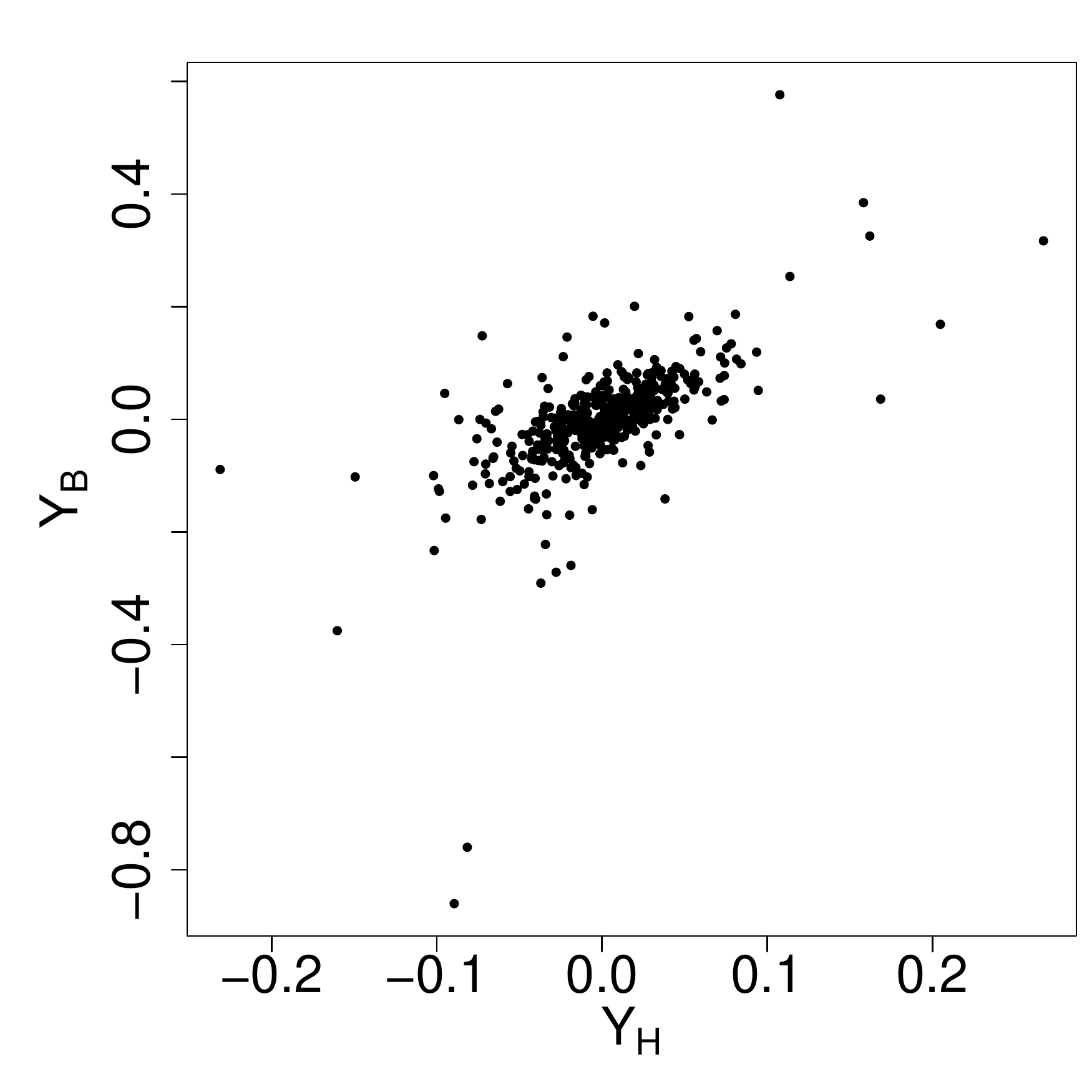}\\
\includegraphics[width=0.24\textwidth]{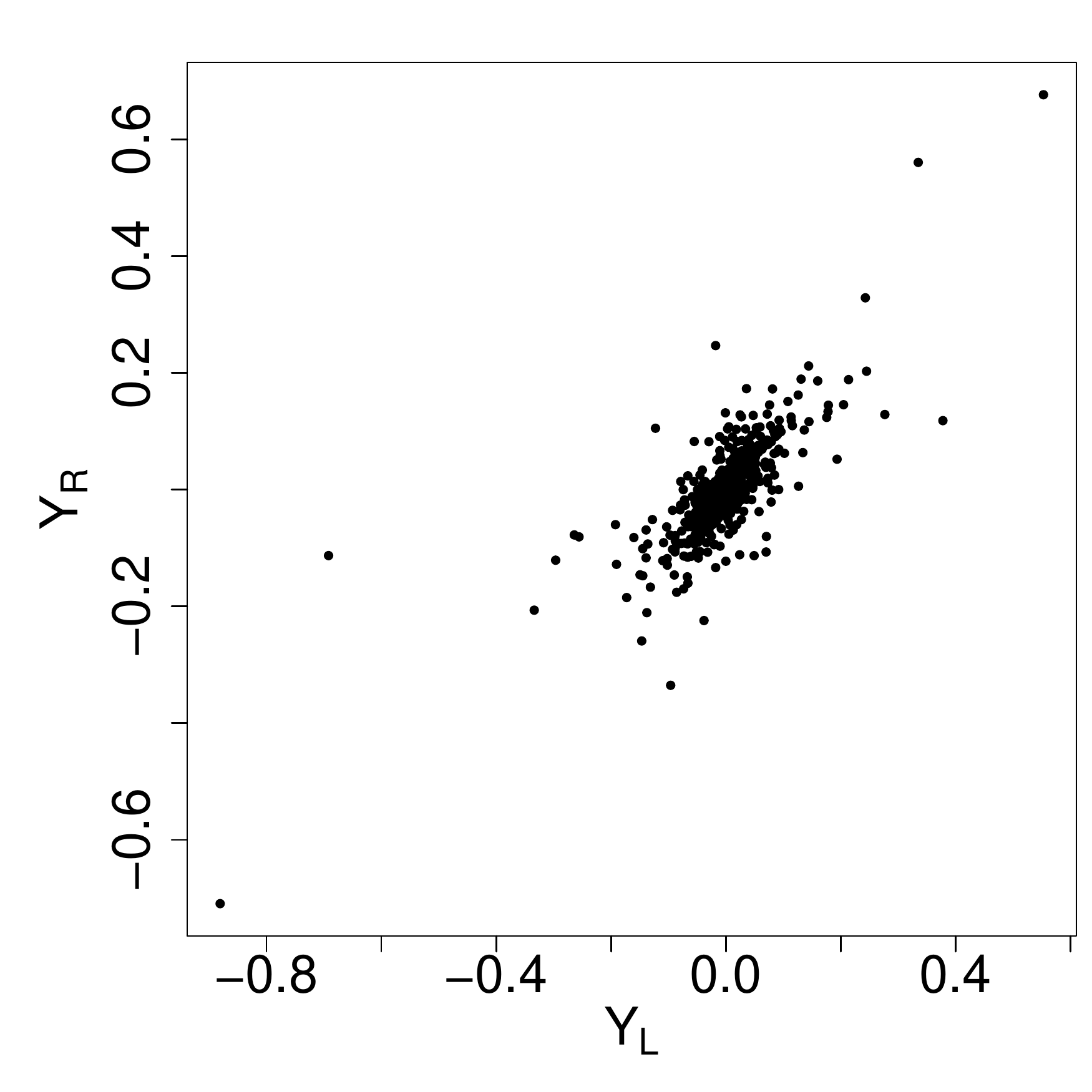}
\includegraphics[width=0.24\textwidth]{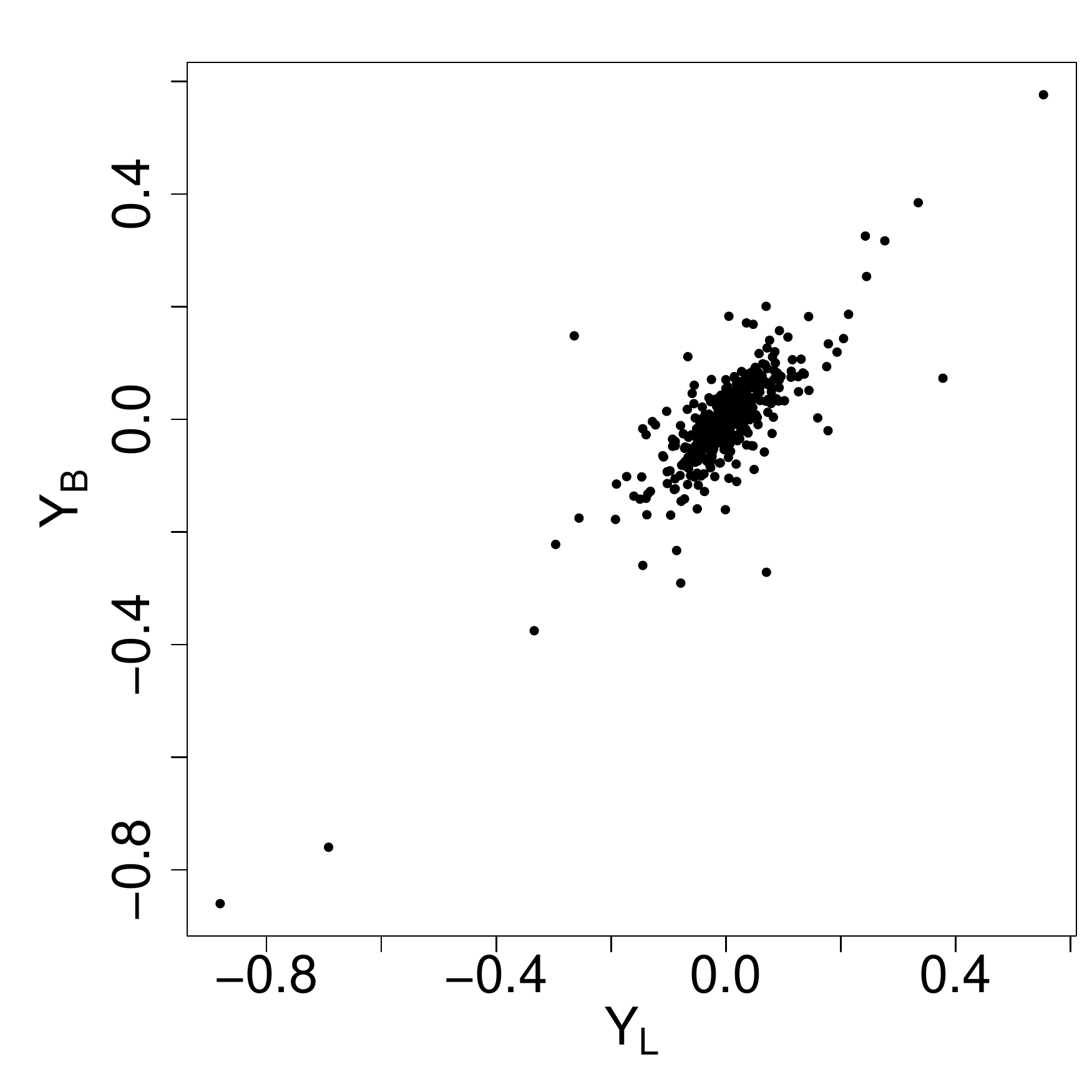}
\includegraphics[width=0.24\textwidth]{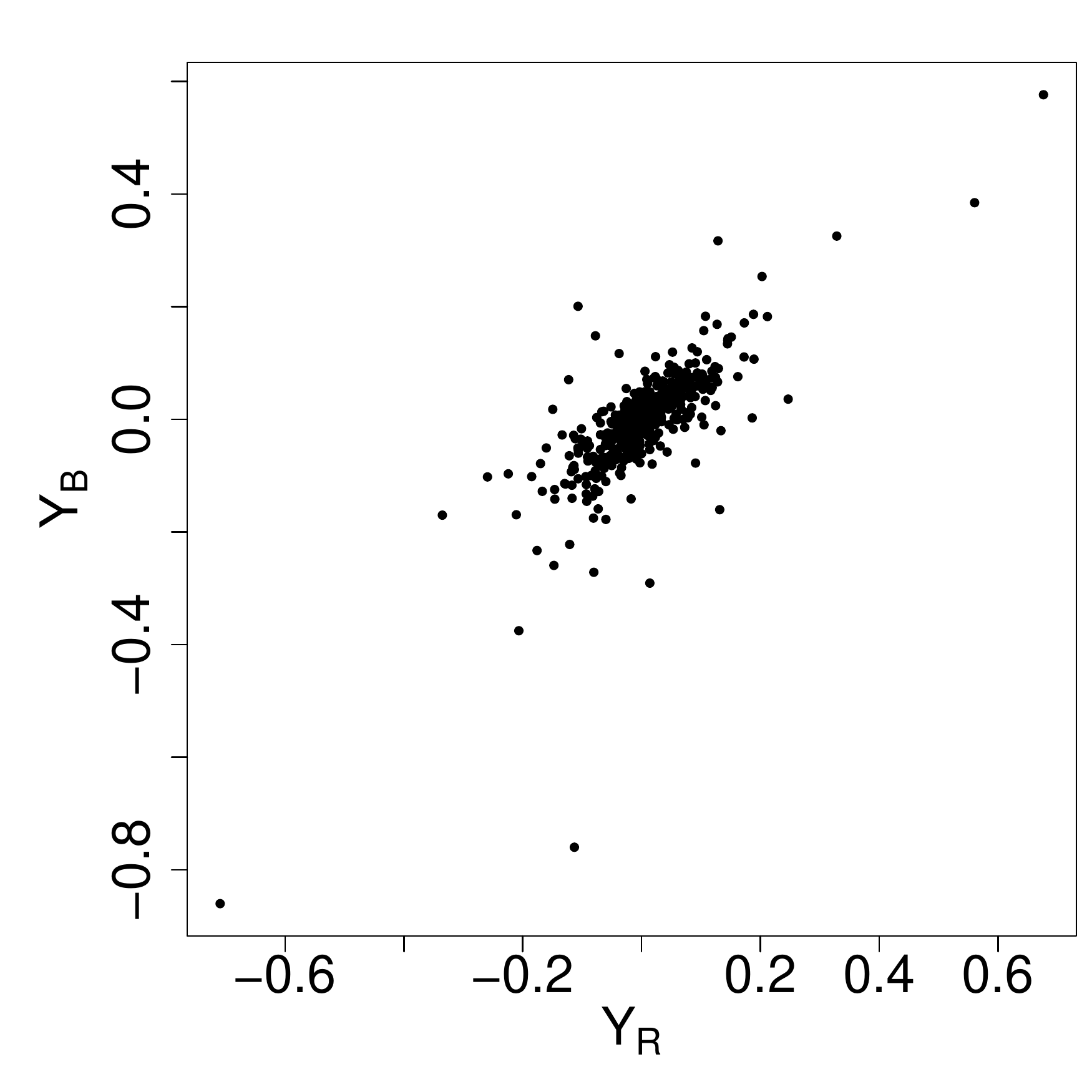}
 \caption{Pairwise scatterplots of the negative weekly returns of the stock prices of four UK banks: HSBC (H), Lloyds (L),  RBS (R) and Barclays (B), from 10/29/2007 to 10/17/2016.}
  \label{fig:bankpw}
\end{figure}

Figure~\ref{fig:bankpw} displays pairwise plots of the negative returns. There is evidence of strong extremal dependence from these plots, as the largest value of $Y_L,Y_R,Y_B$ occurs simultaneously, with positive association amongst other large values. The largest value of $Y_H$ occurs at a different time, but again there is positive association between other large values. As is common in practice the value of $\widehat{\chi}_{HLRB}(q)$ generally decreases as $q$ increases (see Figure~6 in the supplementary material), but is plausibly stable and constant from slightly above $q=0.8$. Consequently, we proceed with fitting a GP distribution.
Ultimately, we wish to fit a parametric GP model to the raw threshold excesses $\{\bm{Y}_t - \bm{u} : \bm{Y}_t \not\leq \bm{u}\}$. In view of the large variety of potential models and parameterizations, we use the model selection strategy detailed in Section~\ref{sec:modelchoice}. Throughout, we use censored likelihood with $\bm{v}=\bm{0}$.

Based on the plot of $\widehat{\chi}_{HLRB}(q)$ we select the 0.83 marginal quantile as the threshold in each margin; there are 149 observations with at least one exceedance. We fit the models with densities \eqref{eq:modconstrgumb}, \eqref{eq:modgumb}, \eqref{eq:modconstrrevexp}, \eqref{eq:modrevexp} and \eqref{eq:modconstrloggauss} to the standardized data. For the final model the matrix $\Sigma$ had diagonal elements fixed at 1, with off-diagonal correlations estimated; this entails some dependence restrictions, see the supplement for further details. The smallest AIC is given by model~\eqref{eq:modconstrgumb}, i.e., where $f_{\bm{T}}$ (see Section~\ref{sec:examples}) is the density of independent Gumbel random variables. We therefore select this class and proceed with item~\eqref{test} of the procedure in Section~\ref{sec:modelchoice} to test for simplifications within this class. In Table~\ref{tab:banktest}, model M1 is the most complex model with all dependence parameters. Model M2 imposes the restriction $\beta_1 = \beta_2 = \beta_3 = \beta_4 = 0$, whilst M3 imposes $\alpha_1 = \alpha_2 = \alpha_3 = \alpha_4 = \alpha$, and M4 imposes both. We observe that both possible sequences of likelihood ratio tests between nested models lead to M4 when adopting a 5\% significance level. This model only contains a single parameter, which is a useful simplification.

\begin{table}[ht]
\centering
\caption{Negative UK bank returns: parameterizations of \eqref{eq:modconstrgumb} for  standardized data.}
\begin{tabular}{llll}
\toprule
 Model & Parameters & Number & Maximized log-likelihood\\
\midrule
 M1 & $\alpha_1,\alpha_2,\alpha_3,\alpha_4, \beta_1, \beta_2, \beta_3$ & 7 & $-917.0$\\
 M2 & $\alpha_1,\alpha_2,\alpha_3,\alpha_4$ & 4 & $-918.2$ \\
 M3 & $\alpha, \beta_1,\beta_2,\beta_3$ & 4 & $-920.8$\\
 M4 & $\alpha$ & 1 & $-921.0$ \\
\bottomrule
\end{tabular}
\label{tab:banktest}
\end{table}

Finally we fit a full GP distribution using Model M4, and test the hypothesis of a common shape parameter. Marginal parameter stability plots suggest that the 0.83 quantile is adequate, which is also supported by diagnostics from the fitted model (supplementary material, Figure~7). At a 5\% significance level, a likelihood ratio test for the hypothesis of $\gamma_H=\gamma_L=\gamma_R=\gamma_B$ provides no evidence to reject the null hypothesis, so a common shape parameter is adopted. The parameter estimates are displayed in Table~\ref{tab:bankfit}.

\begin{table}
\centering
\caption{Negative UK bank returns: maximum likelihood estimates (MLE) and standard errors (SE) of parameters from the final model for the original data.}
 \begin{tabular}{lllllll}
 \toprule
  &$\alpha$& $\sigma_H$ & $\sigma_L$ & $\sigma_R$ & $\sigma_B$  & $\gamma$ \\
 \midrule
  MLE & 1.29 & 0.020 & 0.041 & 0.038 & 0.035 & 0.43\\
  SE & 0.14 & 0.0026 & 0.0053 & 0.0052 & 0.0049 & 0.082\\
 \bottomrule
 \end{tabular}
\label{tab:bankfit}
\end{table}

\begin{figure}[ht]
\centering
 \includegraphics[width=0.24\textwidth]{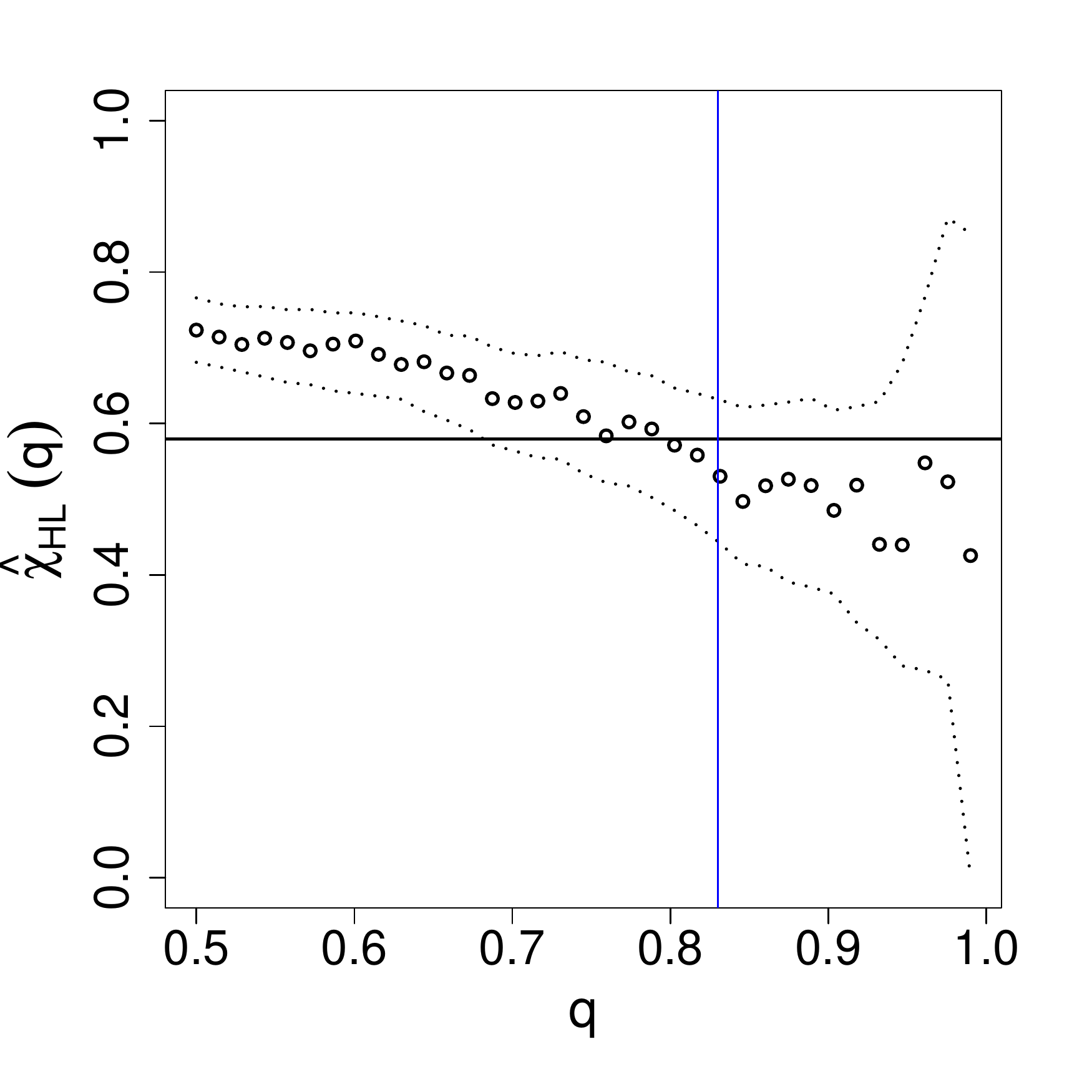}
 \includegraphics[width=0.24\textwidth]{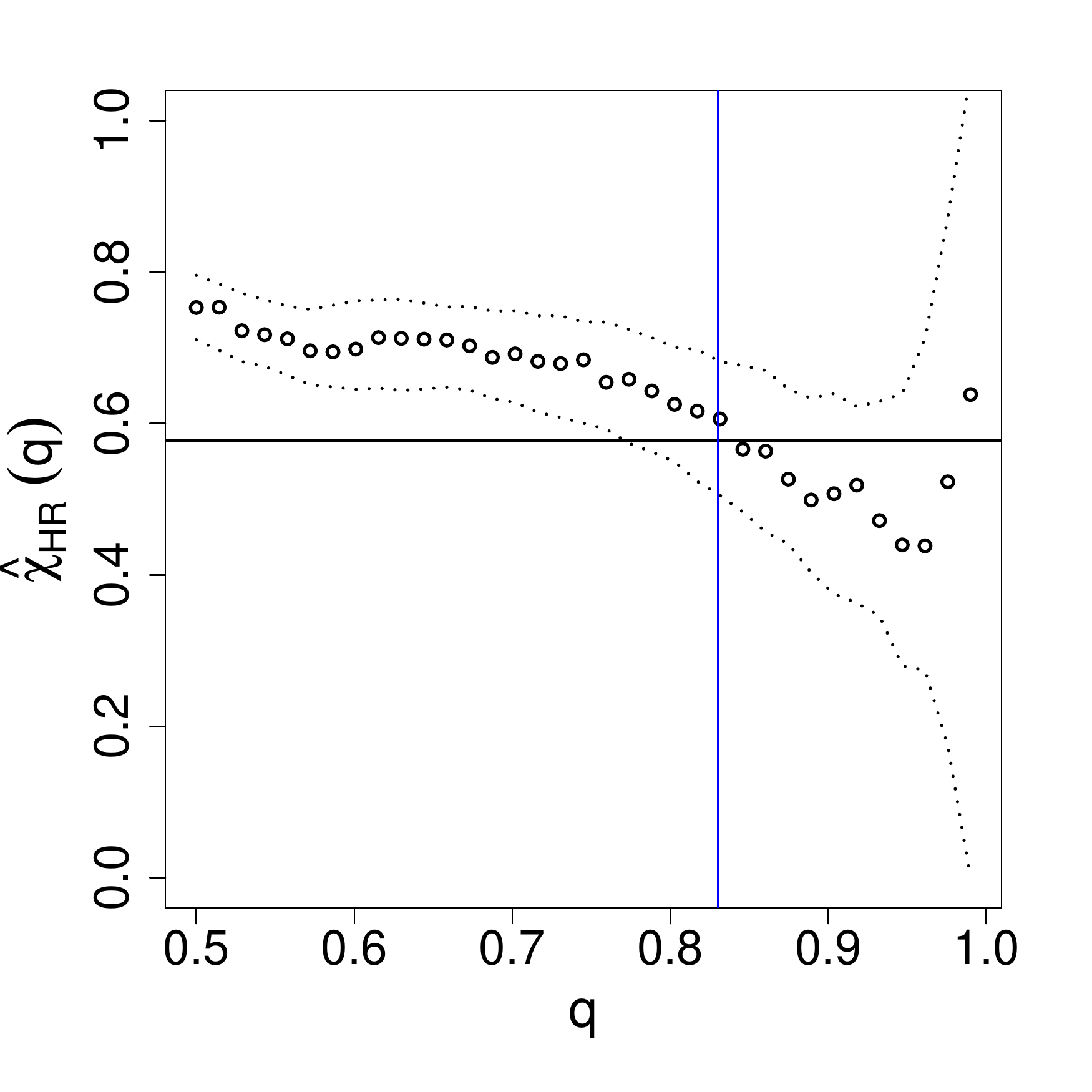}
 \includegraphics[width=0.24\textwidth]{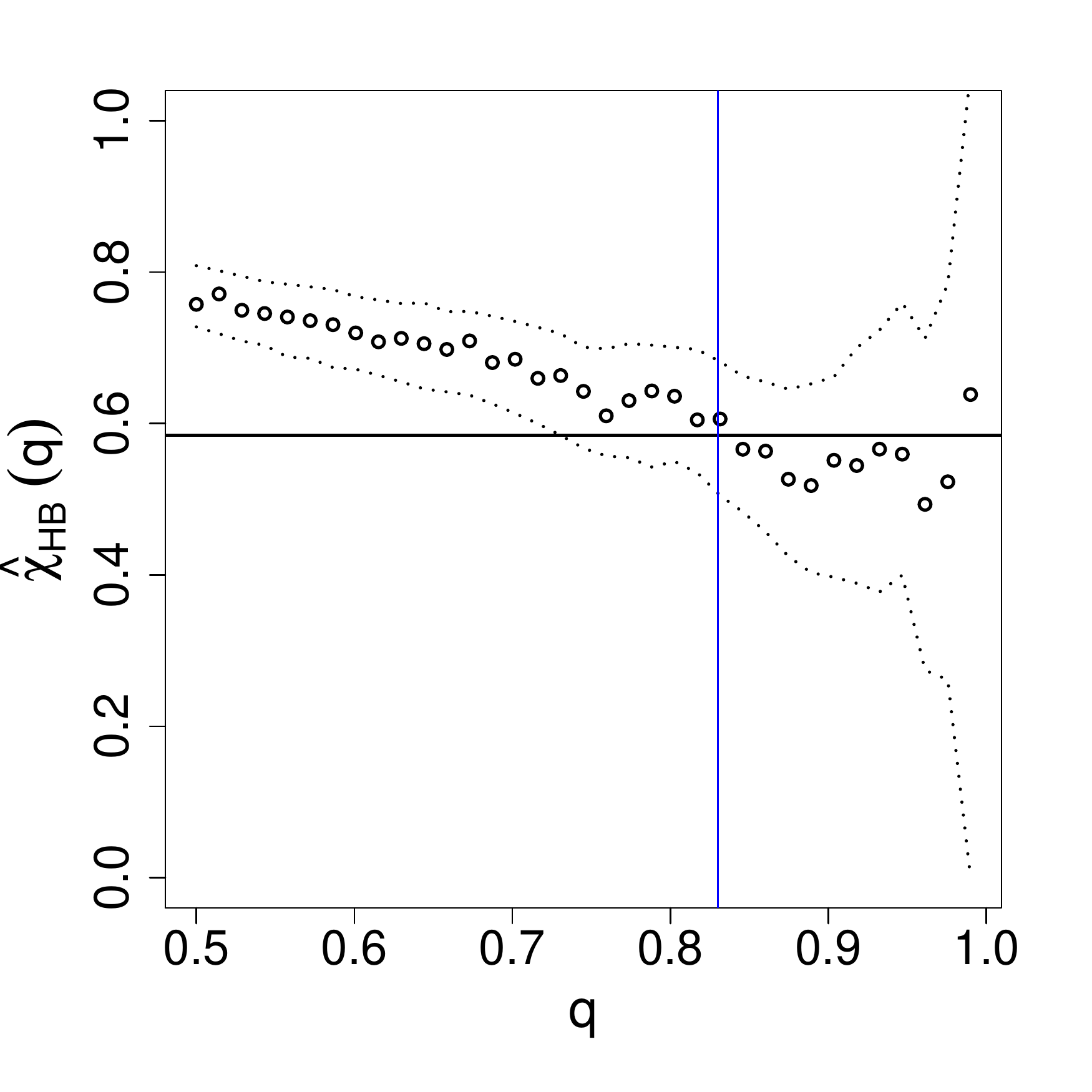}\\
 \includegraphics[width=0.24\textwidth]{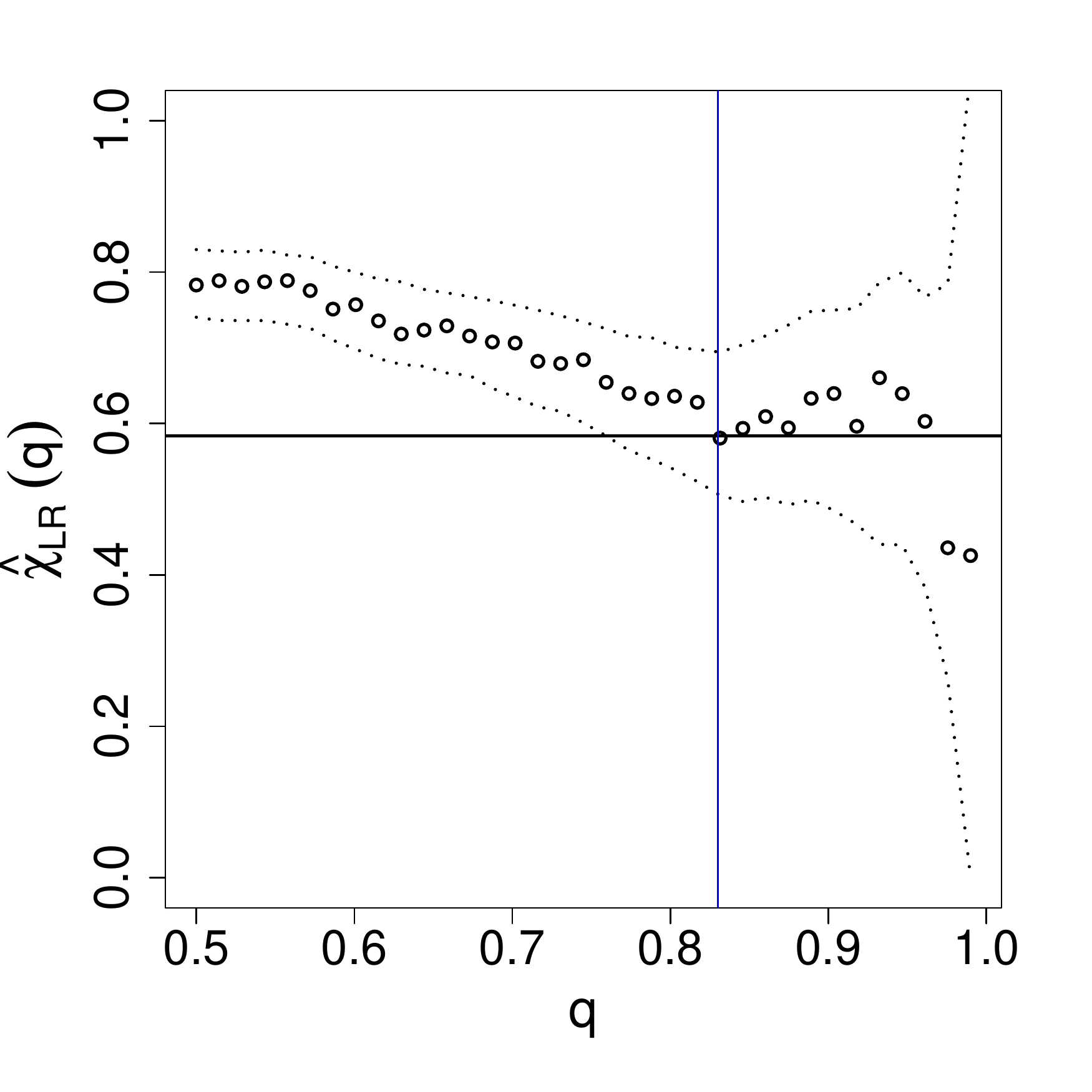}
 \includegraphics[width=0.24\textwidth]{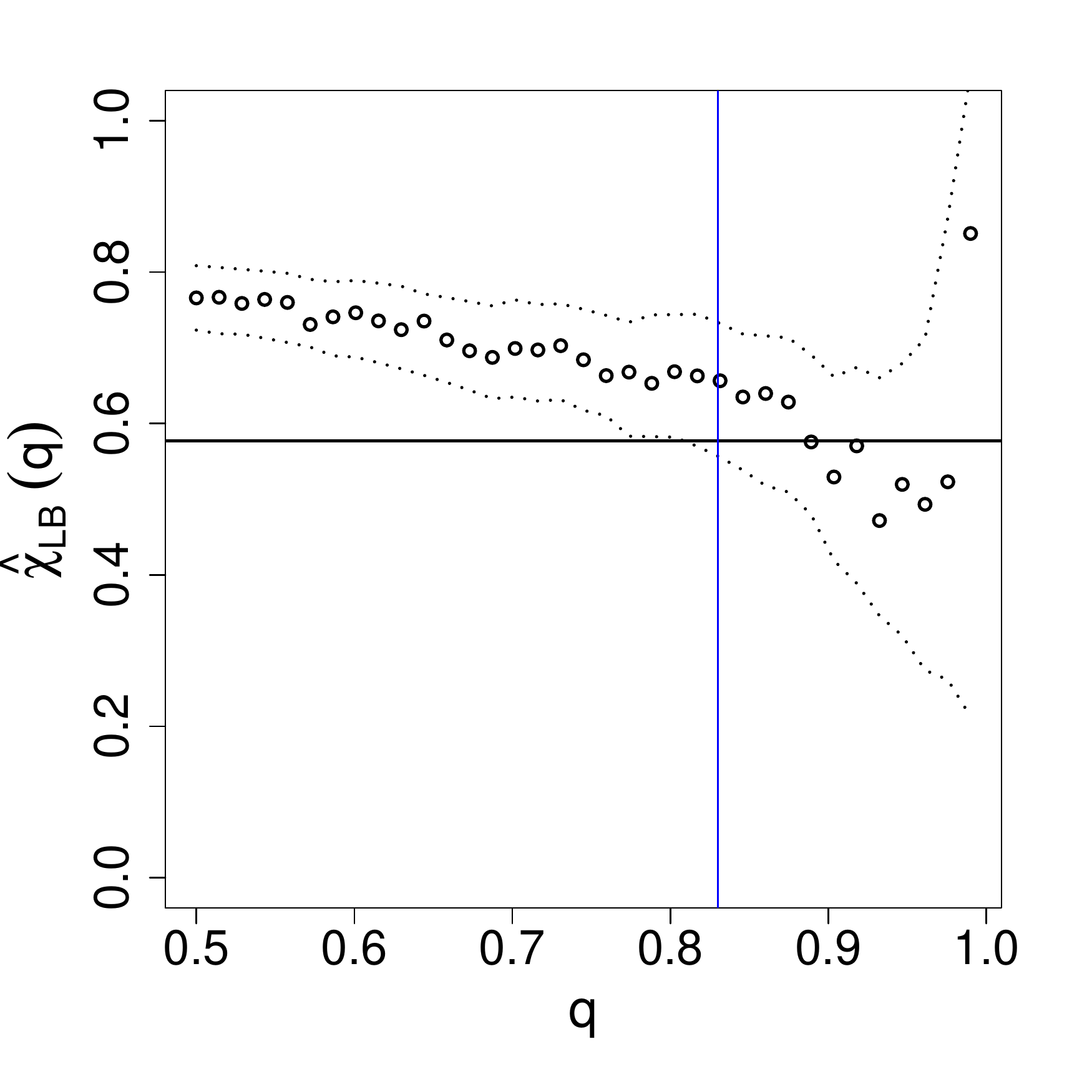}
 \includegraphics[width=0.24\textwidth]{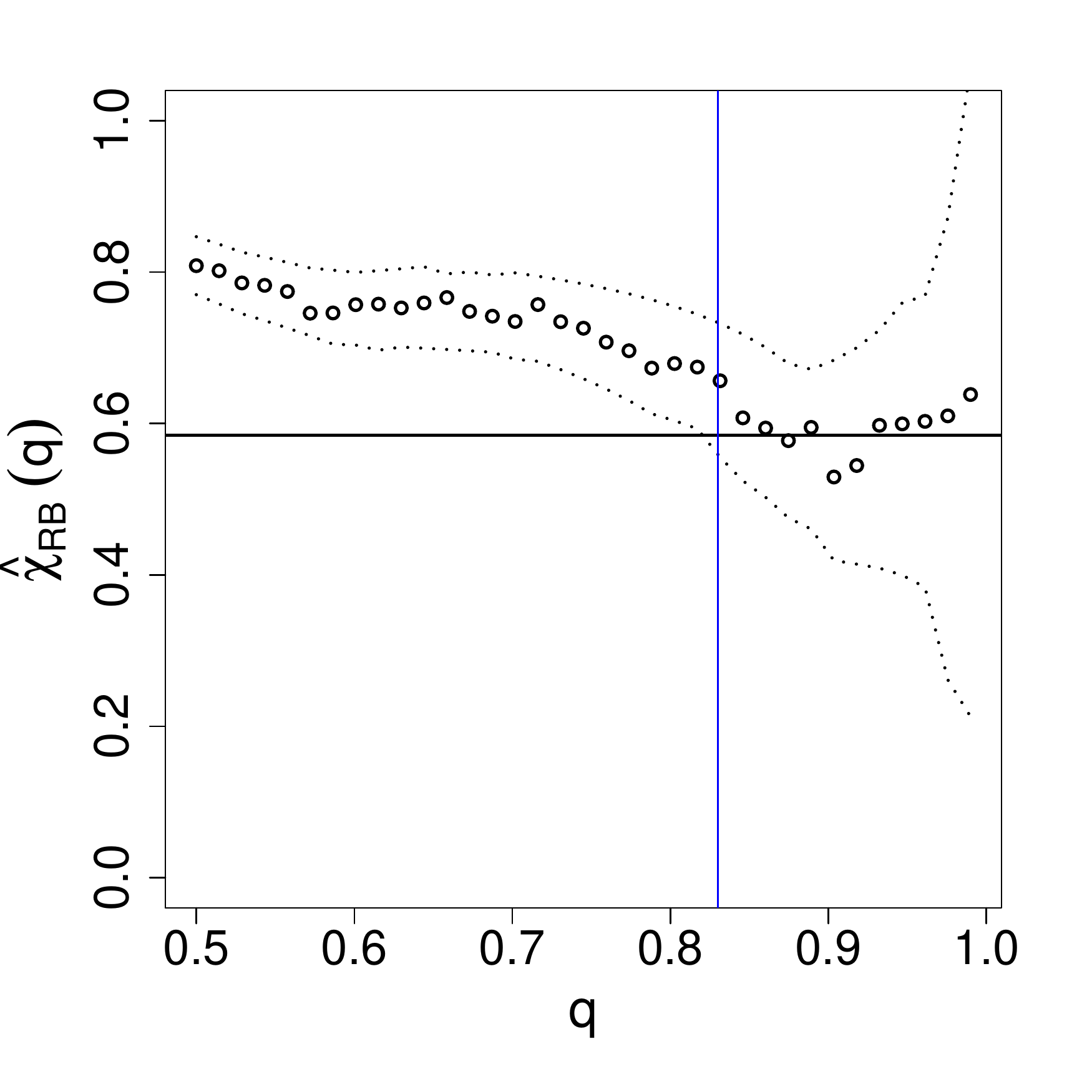}
 \caption{Negative UK bank returns: estimates of pairwise $\chi_{ij}(q)$ with fitted pairwise $\chi_{ij}$ (horizontal line), for HSBC (H), Lloyds (L),  RBS (R) and Barclays (B). Clockwise from top left: $\chi_{HL}$, $\chi_{HR}$, $\chi_{HB}$, $\chi_{RB}$, $\chi_{LB}$, $\chi_{LR}$. The vertical line is the threshold used. Approximate $95 \%$ pointwise confidence intervals are obtained by bootstrapping from $\{\bm{Y}_t:t=1,\ldots,n\}$.}
  \label{fig:bankpwchi}
\end{figure}

To scrutinize the fit of the model, we examine marginal, dependence, and joint diagnostics. Quantile-quantile (QQ) plots for each of the univariate GP distributions implied for $Y_{t,j}-u_j \mid Y_{t,j} >u_j$ are displayed in the supplementary material (Figure~7) indicating reasonable fits in each case. Estimates of the pairwise $\chi_{ij}(q)$, $i\neq j \in\{H,L,R,B\}$, are plotted in Figure~\ref{fig:bankpwchi}, with the corresponding fitted value and threshold indicated; tripletwise plots and the plot of  $\widehat{\chi}_{HLRB}(q)$ show similarly good agreement. Since the model has a single dependence parameter, all pairs are exchangeable and have the same fitted value of $\chi$ for any fixed dimension.

As the shape parameter may be taken as common across margins, we examine the sum-stability property given in~\eqref{eq:sumstab}. We fit a univariate GP distribution to
\begin{align}
\sum_{j\in\{H,L,R,B\}} (Y_{t,j} -u_j) ~\Big| \sum_{j\in\{H,L,R,B\}} (Y_{t,j} -u_j) >0,  \label{eq:possum}
\end{align}
with scale parameter estimate (standard error) obtained as $0.10$ $(0.021)$, and shape parameter estimate $0.45$ $(0.17)$. QQ plots suggest that the fit is good; see the supplementary material (Figure~8). For comparison,  $\sum_{j\in\{H,L,R,B\}} \hat{\sigma}_j = 0.13$ with standard error $0.014$ obtained using the delta method, whilst the maximized univariate GP log-likelihood is $63.5$, and that for the parameters obtained via the multivariate fit is $62.2$, showing that the theory holds well.

Weighted sums of raw stock returns correspond to portfolio performance. We use the final fitted model to compute two commonly-used risk measures, \emph{Value at Risk} (VaR) and \emph{Expected Shortfall} (ES), for a time horizon of one week. If the conditional distribution of $\sum_j a_j (Y_{t,j}-u_j)$ given the event $\sum_j a_j (Y_{t,j}-u_j) > 0$ is $\GP(\sum_j a_j \sigma_j, \gamma)$, then
\begin{align}
 \operatorname{VaR}(p)
 = \sum_j a_j u_j + \frac{\sum_j a_j \sigma_j}{\gamma}\left\{\left(\frac{\phi}{p}\right)^\gamma - 1\right\},
 \label{eq:VaR}
\end{align}
where $0<p<\phi = \Pr[ \sum_j a_j (Y_{t,j}-u_j) > 0]$,  so that~\eqref{eq:VaR} is the unconditional $1-p$ quantile of $\sum_j a_j Y_{t,j}$. We estimate the probability $\phi$  by maximum likelihood using the assumption $\sum_t \I\{\sum_j a_j (Y_{t,j}-u_j) >0\} \sim \operatorname{Bin}(n,\phi)$, and in the univariate model, $\phi$ is orthogonal to the parameters of the conditional excess distribution. In the multivariate model
\begin{equation*}
 \Pr\left[{\textstyle\sum_j} a_j (Y_{t,j}-u_j) > 0\right]
 =
 \Pr\left[{\textstyle\sum_j} a_j (Y_{t,j}-u_j) >0 \mid \bm{Y}_t\not\leq\bm{u}\right]
 \Pr[\bm{Y}_t\not\leq\bm{u}] \notag \\
 =
 p(\bm{\theta}) \, \tilde{\phi},
 \label{eq:phi}
\end{equation*}
where $p(\bm{\theta})$ is an expression involving the parameters of the multivariate GP model, and $\tilde{\phi}$ is the proportion of points for which $\bm{Y}_t\not\leq\bm{u}$. The expression $p(\bm{\theta})$ is not tractable here, thus we continue to estimate $\phi$ as the binomial maximum likelihood estimate, and as a working assumption treat it as orthogonal to the other parameters. However,
an estimate of $p(\bm{\theta})$ can be obtained by simulation using the estimated $\bm{\theta}$; the utility of this will be demonstrated in Figure~\ref{fig:portfolio}.

The expected shortfall is defined as the expected loss given that a particular VaR threshold has been exceeded. Under the GP model, and provided $\gamma < 1$, it is given by
\begin{equation*}
  \operatorname{ES}(p)
  =
  \textstyle \E \left[ \sum_j a_j Y_{t,j} \mid \sum_j a_j Y_{t,j} > \operatorname{VaR}(p) \right] \notag \\
  =
  \operatorname{VaR}(p) + \frac{\sum_j a_j \sigma_j +\gamma \left[\operatorname{VaR}(p)-\sum_j a_j u_j\right]}{1-\gamma}. \label{eq:ES}
\end{equation*}
Asymptotic theory suggests that a univariate GP model fit directly to $\sum_{j}a_j(Y_{t,j}-u_j)$ or the implied GP$(\sum_j a_j \sigma_j , \gamma)$ model obtained from the multivariate fit could be used. An advantage of using the GP$(\sum_j a_j \sigma_j , \gamma)$ model derived from the multivariate fit is reduced uncertainty, combined with consistent estimates across different portfolio combinations.

\begin{figure}[ht]
\centering
 \includegraphics[width=0.24\textwidth]{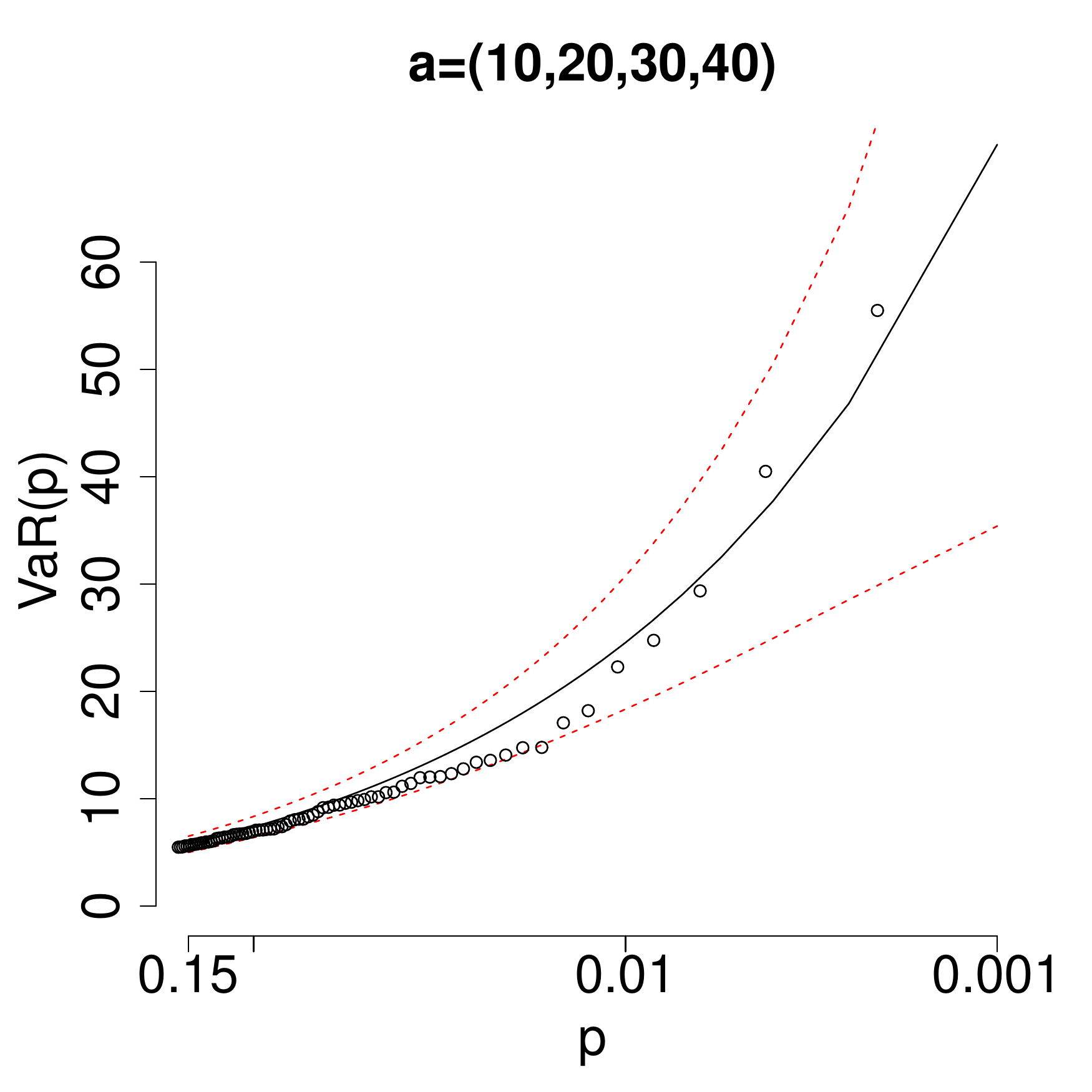}
 \includegraphics[width=0.24\textwidth]{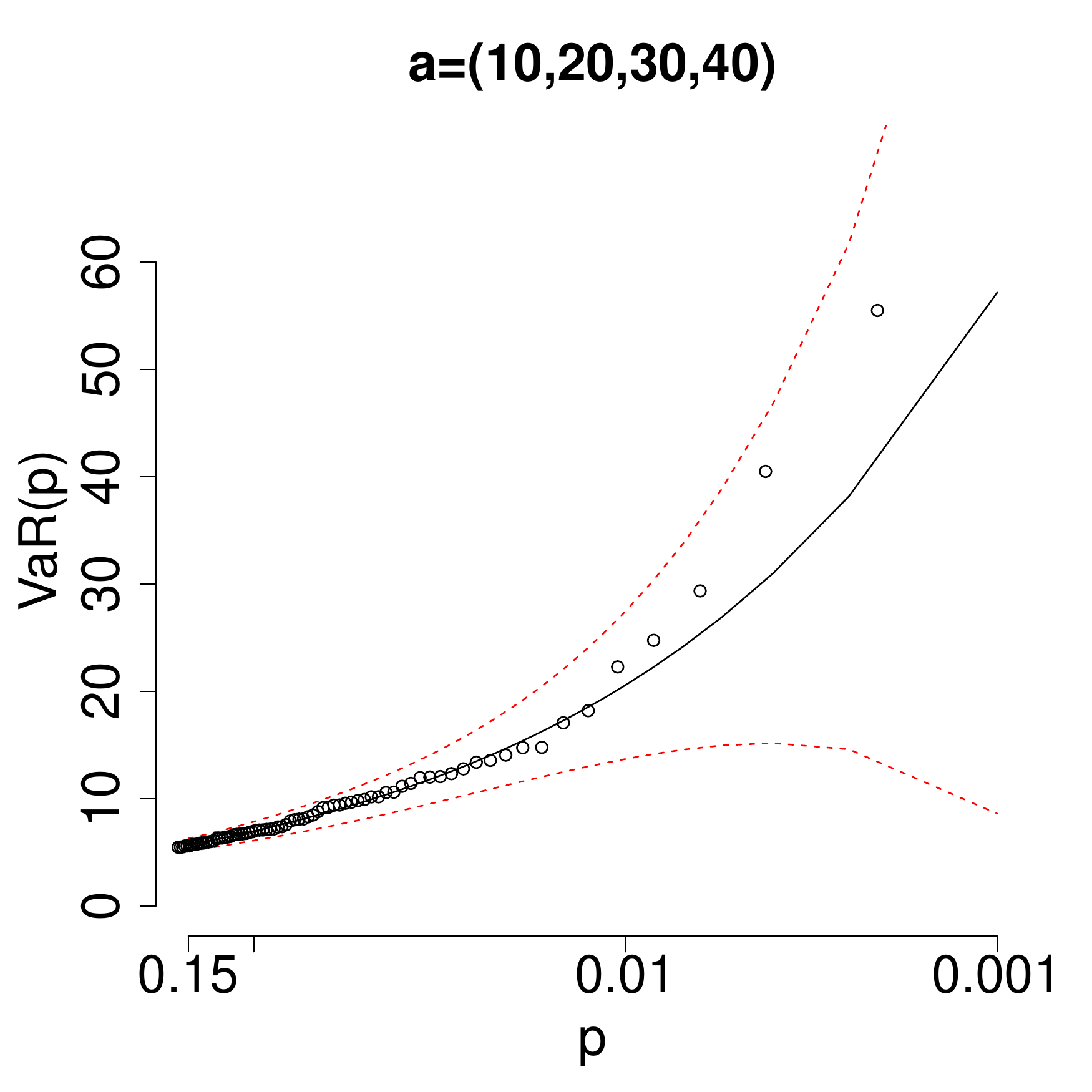}
  \includegraphics[width=0.24\textwidth]{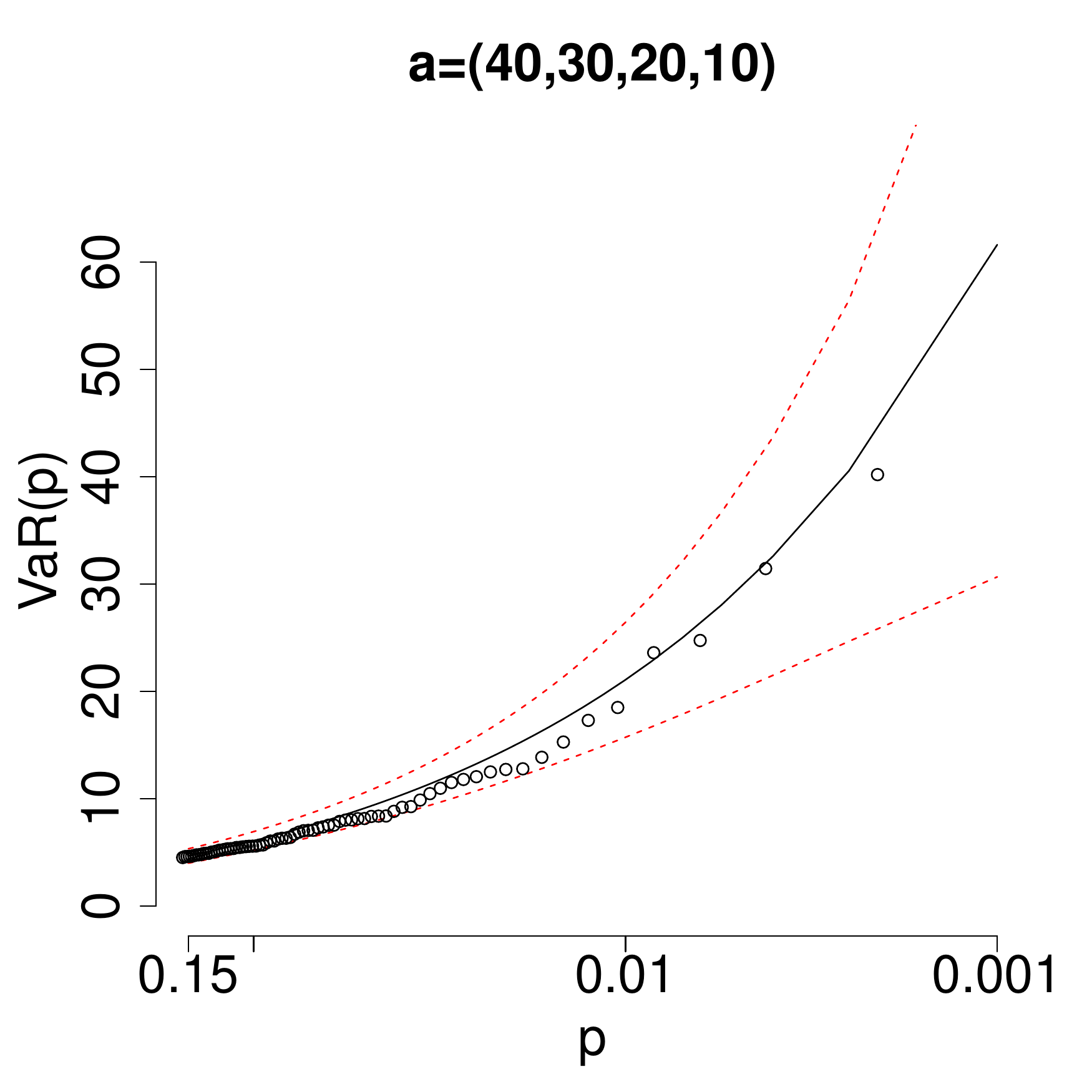}
 \includegraphics[width=0.24\textwidth]{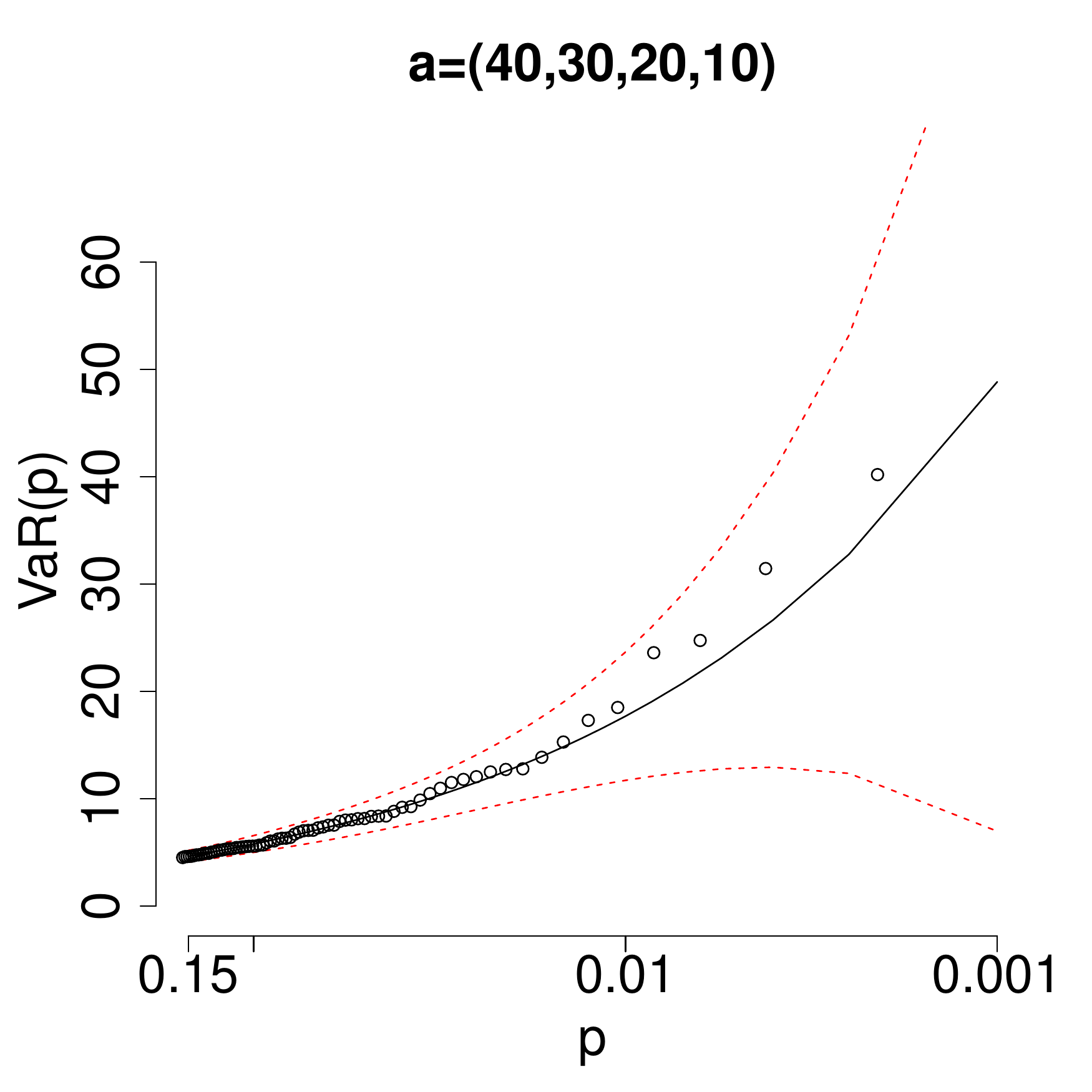}
 \caption{VaR estimates and pointwise 95\% delta-method confidence intervals for portfolio losses based on the weights given as percentages invested in HSBC, Lloyds,  RBS and Barclays as in the figure title. Estimates based on the multivariate GP fit are on the left of a pair; estimates based on the univariate fit are on the right.}
  \label{fig:var}
\end{figure}

\begin{figure}[ht]
\centering
 \includegraphics[width=0.3\textwidth]{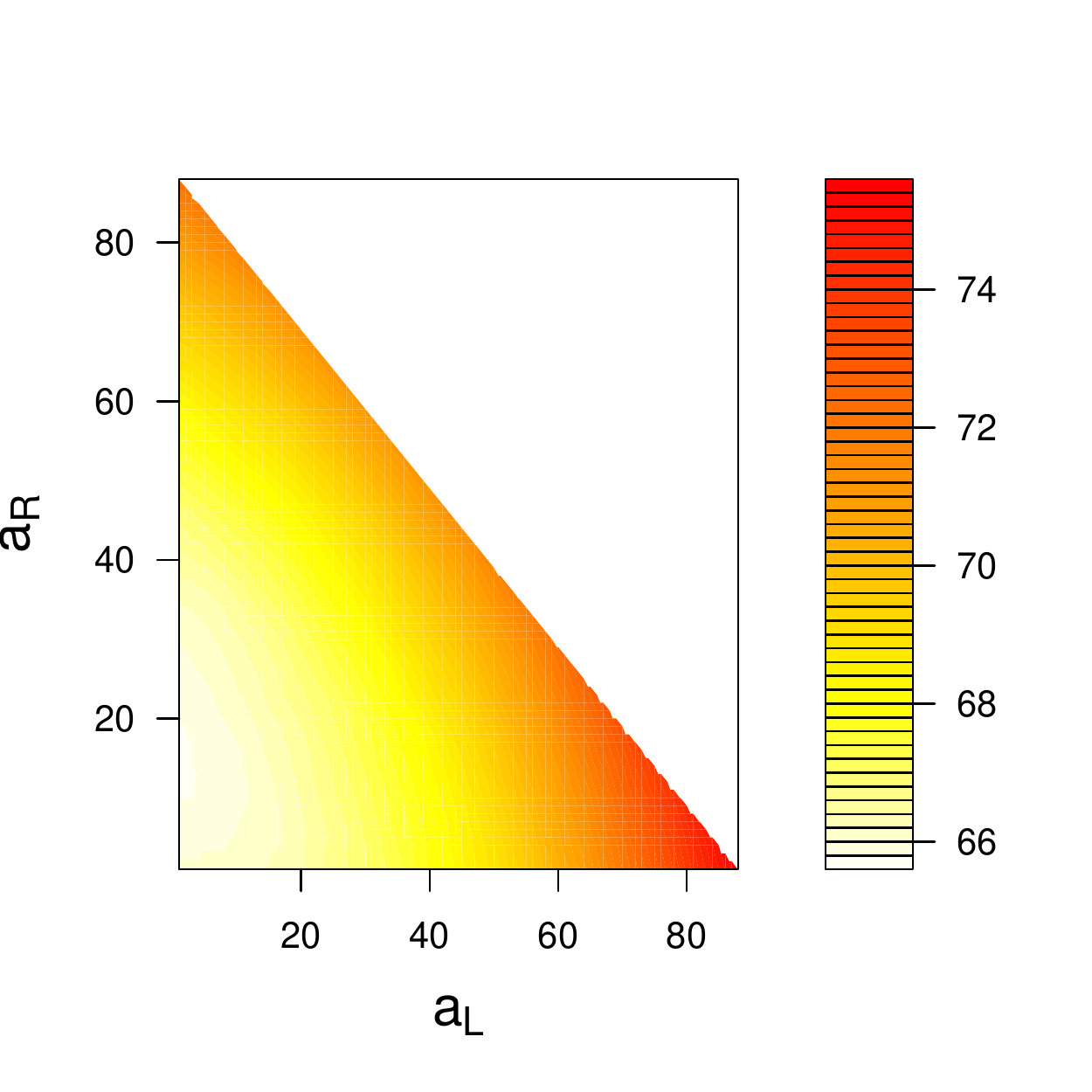}
 \includegraphics[width=0.3\textwidth]{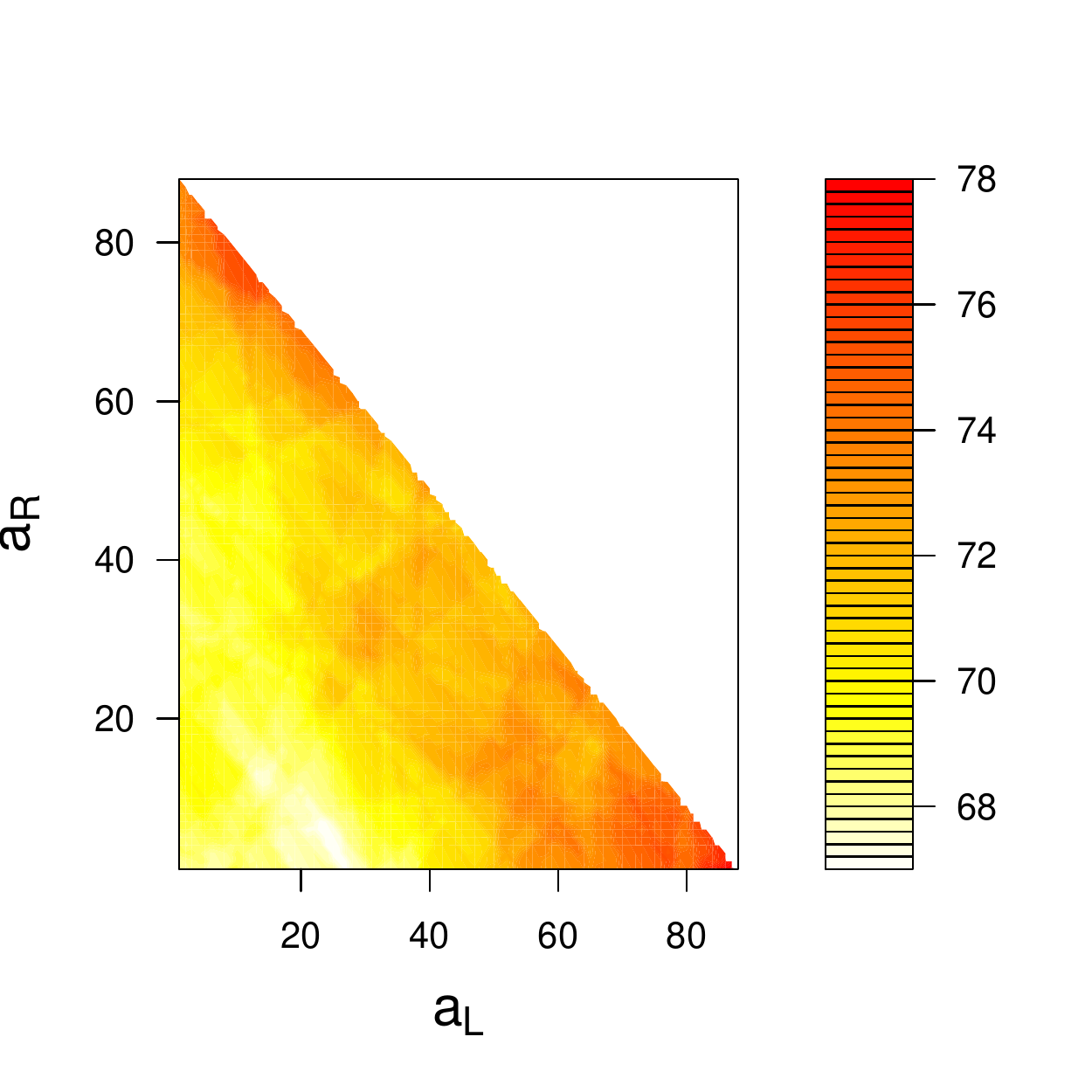}
  \includegraphics[width=0.3\textwidth]{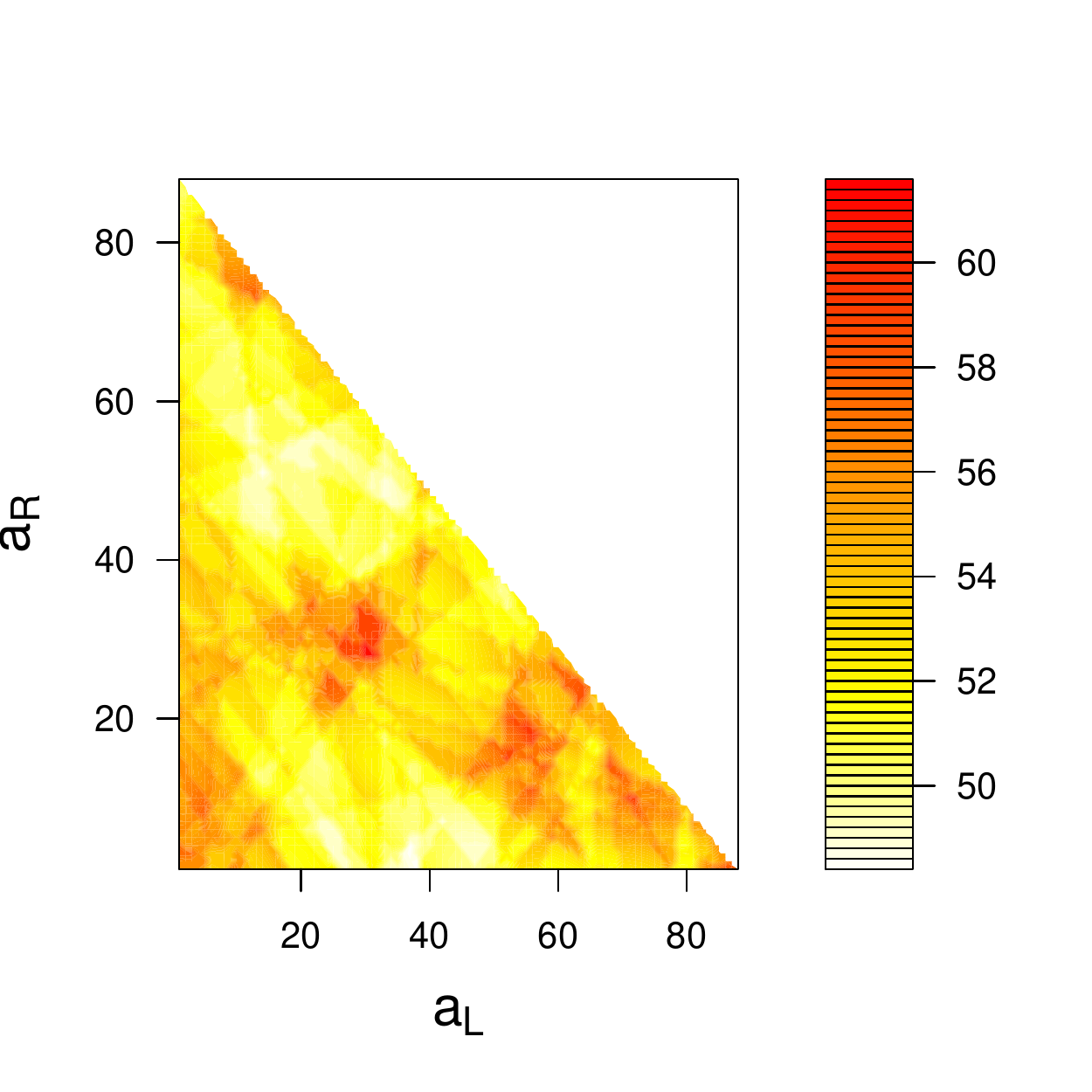}
 \caption{Maximum likelihood estimates of VaR(0.001) for $\sum_j a_j Y_{t,j}$ with $a_H=10$ and $a_B=90 - a_L - a_R$ representing a portfolio of stocks of HSBC, Lloyds,  RBS and Barclays. Left: from multivariate model including simulation to estimate $p(\bm{\theta})$ from~\eqref{eq:phi}; centre: from multivariate model using the binomial estimate of $\phi$; right: from univariate model fit to each combination separately. Note the different colour scales on each panel.}
  \label{fig:portfolio}
\end{figure}

Figures~\ref{fig:var} displays VaR curves and confidence intervals for two different weight combinations and for both the univariate and multivariate fits, together with empirical counterparts, whilst Figure~9 in the supplementary material shows the corresponding ES curves. For VaR the univariate fit is closer in the body and the multivariate fit is closer to the data in the tails. The reduction in uncertainty is clear and potentially quite useful for smaller $p$. For ES (supplementary material, Figure~9) the univariate fit estimates smaller values than the multivariate fit in each case, and seems to reflect the observed data better. However, the empirical ES values fall within the 95\%  confidence intervals obtained from the multivariate model, suggesting that the model is still consistent with the data. Note that the univariate fit is tailored specifically to the data $\sum_j a_j Y_{t,j}$ and as such, we would always expect the point estimates from Figure~\ref{fig:var} to look better for the univariate fit. On the other hand, when interest lies in different functions of the extremes of $Y_{t,j}$, the multivariate approach is able to deliver self-consistent inference.

Figure~\ref{fig:portfolio} illustrates how the multivariate model provides more consistent estimates of VaR across different portfolio combinations compared to the use of multiple univariate models.
To produce the figures, we suppose that $\sum_j a_j = 100$ represents the total amount available to invest. The value $a_H=10$ is fixed, with other weights varying, but with each $a_j\geq 1$. Two estimates making use of the multivariate model are provided: one for which a model-based estimate of $p(\bm{\theta})$ from~\eqref{eq:phi} is used (with estimation based on 100\,000 draws from the fitted model), and one where the empirical binomial estimate of $\phi$ is used, as in Figure~\ref{fig:var} and the supplementary material (Figure 9). Both sets of multivariate estimates suggest much more consistent behaviour across portfolio combinations than the use of univariate fits. In particular, behaviour is very smooth once a model-based estimate for $p(\bm{\theta})$ is included.

\section{Landslides}
\label{sec:rainfall}

Rainfall can cause ground water pressure build-up which, if very high, can trigger a landslide. The cause can be short periods with extreme rain intensities, or longer periods of up to three days of more moderate, but still high rain intensities. \citet{guzzetti2007} consolidate many previous studies and propose threshold functions which link duration in hours, $D$, with total rainfall in millimeters, $P$, such that rainfall below these thresholds are unlikely to cause landslides. For highland climates in Europe this function is
\begin{equation}\label{eq:ID}
  P=7.56 \times D^{0.52}.
\end{equation}
Thus, a one-day rainfall below 39.5 mm, a two-day rainfall below 56.6 mm, or a three-day rainfall below 69.9 mm are all unlikely to cause a landslide.

We  use a long time series of daily precipitation amounts $P_1,\ldots,P_N$ collected by the Abisko Scientific Research Station in northern Sweden in the period 1/1/1913 -- 12/ 31/2014, to  estimate a lower bound for the probability of the occurrence of rainfall events  which may  lead to landslides.
The total cost of landslides in Sweden is  around SEK 200 million/year. There have been several landslides in the Abisko area in the past century, for instance in October 1959, August 1998, and July 2004 \citep{rapp1976,jonasson1999,beylich2005}. The rainfall episodes causing the landslides are clearly visible in the data, with 24.5 mm of rain
on October 5, 1959, 21.0 mm of rain on August 24, 1998, and 61.9 mm of rain on July 21, 2004. The 2004 rain amount is well above the 1-day risk threshold, whereas the 1959 and 1998 rain amounts are below the 1-day threshold. The explanation may be that the durations of the latter two rain events were shorter than 24 hours, and that the threshold in~\eqref{eq:ID} was still exceeded.

\begin{figure}[ht]
\centering
\includegraphics[width=0.6\textwidth]{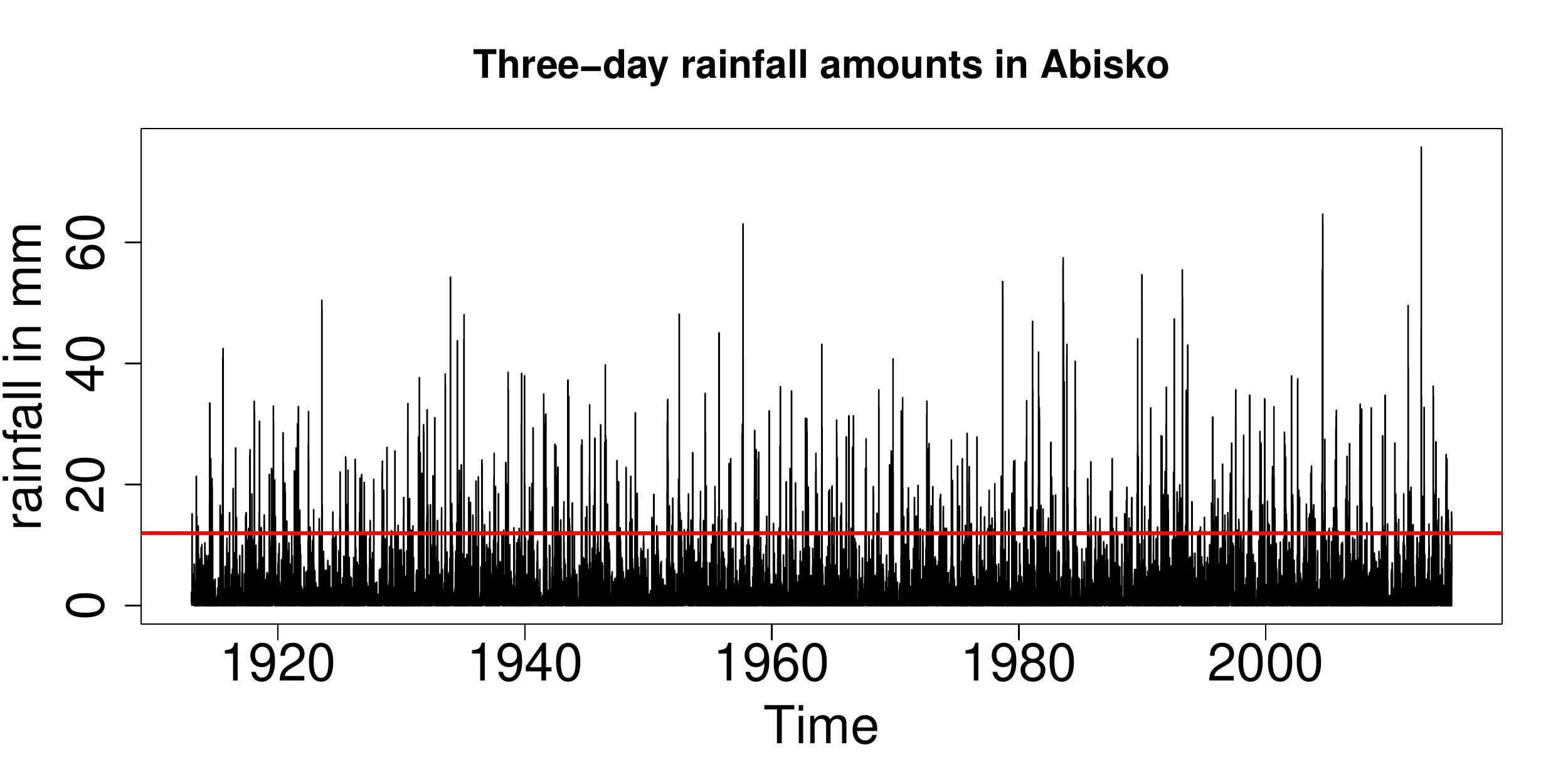}
\caption{Precipitation data in Abisko: cumulative three-day precipitation amounts  $P_i + P_{i+1} + P_{i+2}$ for $i \in \{1,\ldots, N-2\}$ with threshold $u = 12$ in red.}
\label{fig:landslide1}
\end{figure}

We wish to construct a dataset $\bm{Y}_1,\ldots,\bm{Y}_n \in \reals^3$, for $n < N$, whose components represent daily, two-day, and three-day extreme rainfall amounts respectively, to account for longer periods of moderate rainfall. Based on a mean residual life plot and parameter stability plots (not shown here) for the daily rainfall amounts $P_1,\ldots,P_N$, we choose the threshold $u = 12$, which corresponds roughly to the $99 \%$ quantile. Figure~\ref{fig:landslide1} shows the cumulative three-day precipitation amounts $P_i + P_{i+1} + P_{i+2}$ for $i \in \{1,\ldots, N-2\}$. The threshold $u$ is used to extract clusters of data containing extreme episodes; the data $\bm{Y}_1,\ldots,\bm{Y}_n$ are constructed as follows:

\begin{compactenum}
\item Let $i$ correspond to the first sum $P_i + P_{i+1} + P_{i+2}$ which exceeds the threshold $u$ and set $P_{(1)} = \max(P_i,P_{i+1},P_{i+2})$.

\item Let the first cluster $C_{(1)}$ consist of $P_{(1)}$ plus the five values preceding it and the five values following it.

\item Let $Y_{11}$ be the largest value in $C_{(1)}$, $Y_{12}$ the largest sum of two consecutive non-zero values in $C_{(1)}$, and $Y_{13}$ the largest sum of three consecutive non-zero values in  $C_{(1)}$.

\item Find the second cluster $C_{(2)}$ and compute $\bm{Y}_2 = (Y_{21}, Y_{22}, Y_{23})$ in the same way, starting with the first observation after $C_{(1)}$.
\end{compactenum}
Continuing this way, we obtain a dataset $\bm{Y}_1,\ldots,\bm{Y}_n$, with $d=3$  and $n = 580$.

Annual maxima of a similar data set were analysed in \citet{rudvik2012}, with the conclusion that there was no time trend. We fitted a univariate GP distribution with a fixed shape parameter $\gamma$ but a loglinear trend for the scale parameter to the marginal components  $(\bm{Y}_{i})_{i=1}^n$, and also did not find any significant trend; see the supplementary material. The estimated shape parameters obtained from fitting univariate GP distributions to the marginal threshold excesses are close to zero (the hypothesis $\bm{\gamma} = \bm{0}$ is not rejected at a $5 \%$ level) and the confidence intervals for the scale parameters overlap (Table~\ref{tab:margins}). Note that a common $\sigma$ and $\gamma$ only implies that the marginal distributions are equal \emph{conditional on} exceeding the threshold; it does not imply that the unconditional probabilities $\mathbb{P}[Y_j > u_j]$ are equal.

\begin{table}[ht]
\centering
 \caption{Precipitation data in Abisko: estimates of the parameters of marginal GP models for thresholds $u = 12$, $u = 13.5$ and $u = 14$ respectively; standard errors in parentheses.}
\begin{tabular}{lccccc}
\toprule
  & $\bm{Y}_{i1}$ & $\bm{Y}_{i2}$ & $\bm{Y}_{i3}$ \\
\midrule
 $\widehat{\gamma}$ & -0.06 (0.05) & -0.02 (0.06) & -0.01 (0.05) \\
 $\widehat{\sigma}$ & \phantom{-}8.26 (0.69) & \phantom{-}9.34 (0.74) & \phantom{-}9.96 (0.74) \\
\bottomrule
\end{tabular}
  \label{tab:margins}
\end{table}

In the following analysis, we  set $\bm{\sigma} = \sigma \bm{1}$ and $\bm{\gamma} = \gamma \bm{1}$, and we fit the structured models from Section~\ref{sec:structured}, both with $\gamma = 0$ and with $\gamma > 0$, using censored likelihood with $\bm{v} = \bm{0}$. To ensure identifiability we set $\lambda_1 = 1$ for both models. We choose $\bm{u} = u \bm{1}$ with $u= 24$ since parameter estimates stabilize for thresholds around this value, and continue with the $142$ data points whose third components exceed $u = 24$.

The estimates of $\sigma$ are somewhat higher than in the marginal analysis and again the hypothesis $\gamma = 0$  was not rejected (Table~\ref{tab:fit}). The higher estimate of $\sigma$ is intuitively reasonable since the maximum likelihood estimators for $\gamma$ and $\sigma$ are negatively correlated and since $\widehat{\gamma}$ is positive for the second model.

To estimate the risk of a future landslide we assume that the extreme rainfalls, i.e., the $142$ data points whose third components exceed $u = 24$, occur in time as a Poisson process.  The number of extreme rainfalls in a year then follows a Poisson distribution whose mean we will denote by $\zeta$. Assuming that the sizes of the excesses are independent of the Poisson process, the yearly number of rainfalls for which at least one component exceeds  the risk level $\bm{y} = (39.5,56.6,69.9)$ (obtained from  \eqref{eq:ID}) has a Poisson distribution with parameter
\begin{equation}\label{eq:approx2}
\mu =  \zeta \left\{ 1 - H \left( \frac{\bm{y}- \bm{u}}{\bm{\sigma}} ; \bm{1},\bm{0} \right) \right\}.
\end{equation}
Estimating $\zeta$ by $\frac{\#\text{extreme rainfalls}}{\#\text{years}}=142/102$ and $H$ by integrating the density~\eqref{eq:orderdens}, using the parameter estimates $(\widehat{\lambda}_1,\widehat{\lambda}_2, \widehat{\lambda}_3, \widehat{\sigma})$
from the top row of Table~\ref{tab:fit}, we obtain the estimate $\widehat{\mu} = 0.102$. Hence, for any given year, the probability that there is exactly one rainfall episode which could lead to a landslide is $0.092$, and the probability that there is at least one such rainfall is $0.097$. This is higher than the result in \citet{rudvik2012} who used data from 1913--2008 and analysed daily, three-day and five-day precipitation amounts to estimate the yearly risk of at least one dangerous rainfall episode. In the data, we observed seven exceedances of $\bm{y}$ over 102 years. This is not too far from the ten extreme rainfalls that we would expect based on our model.


\begin{table}[ht]
\centering
 \caption{Precipitation data in Abisko: parameter estimates for the structured components model with $u = 24$; standard errors in parentheses.}
\begin{tabular}{lcccccc}
\toprule
Model  & $\widehat{\lambda}_1$ & $\widehat{\lambda}_2$ & $\widehat{\lambda}_3$ & $\widehat{\sigma}$ & $\widehat{\gamma}$ &  Log-likelihood \\
\midrule
 $\gamma = 0$ & 1.00 & 0.84 (0.13) & 1.08  (0.18) & 10.17  (0.80) & 0 & -870.0 \\
 $\gamma > 0$ & 1.00 &  0.83 (0.12) & 1.06 (0.18) & 9.14  (0.99)  & 0.11 (0.08) & -868.9 \\
\bottomrule
\end{tabular}
  \label{tab:fit}
\end{table}

Marginal QQ-plots show good fits for components~2 and~3, but less so for component~1 for the model with  $\gamma = 0$ (Figure~5 in the supplementary material). This is due to the restriction $\bm{\sigma} = \sigma \bm{1}$  used to ensure that the components are ordered.

\begin{figure}[ht]
\centering
\subfloat{\includegraphics[width=0.3\textwidth]{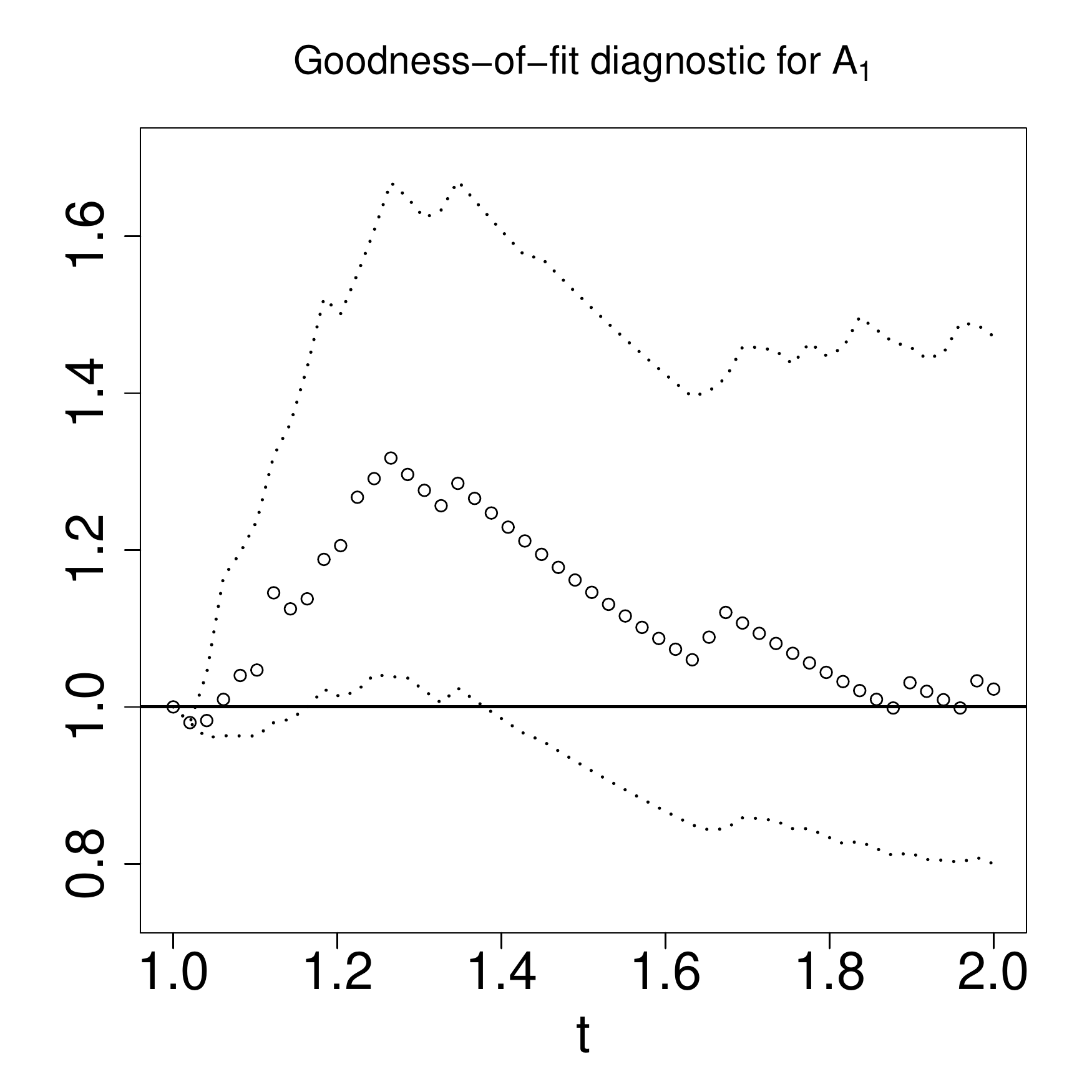}}
\subfloat{\includegraphics[width=0.3\textwidth]{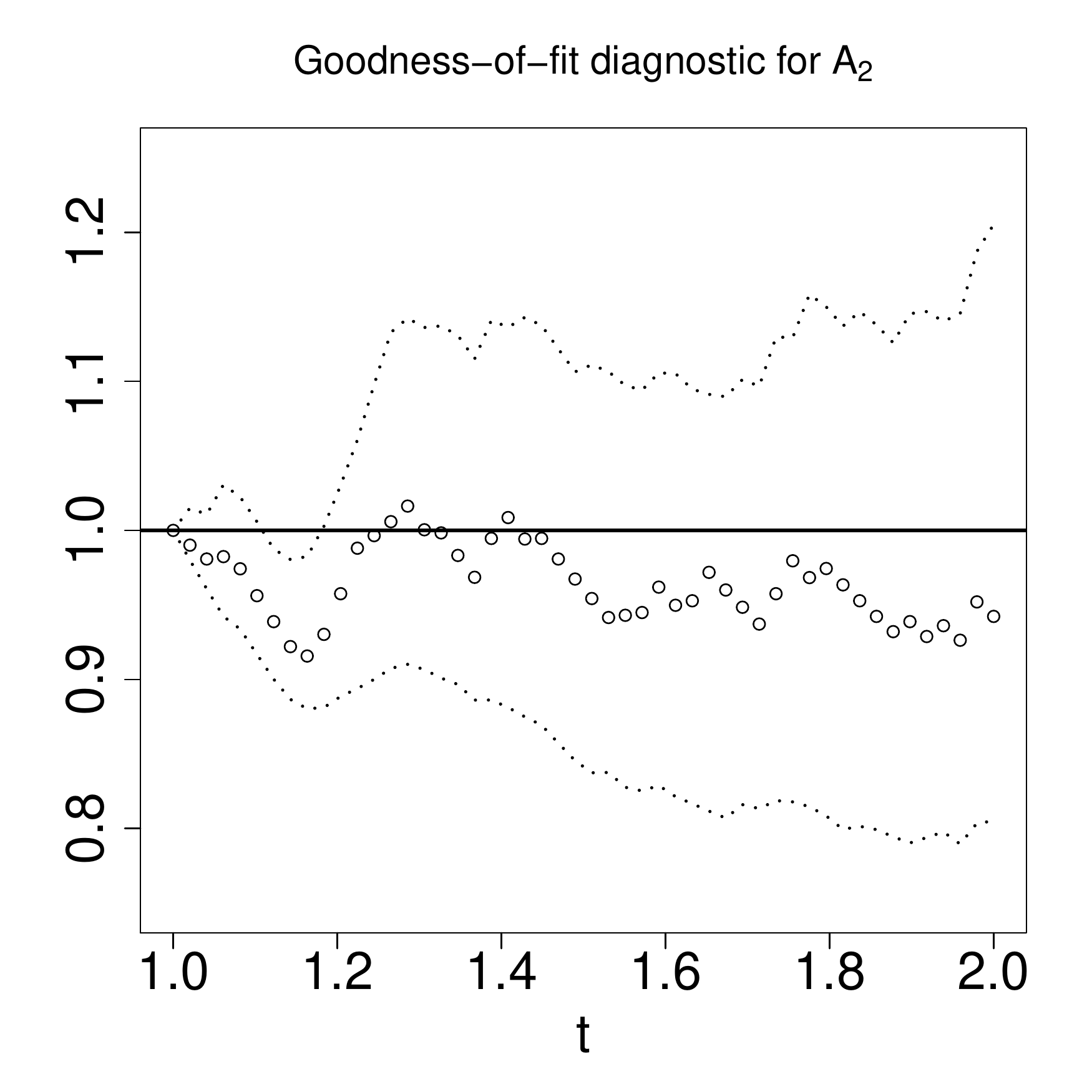}}
\subfloat{\includegraphics[width=0.3\textwidth]{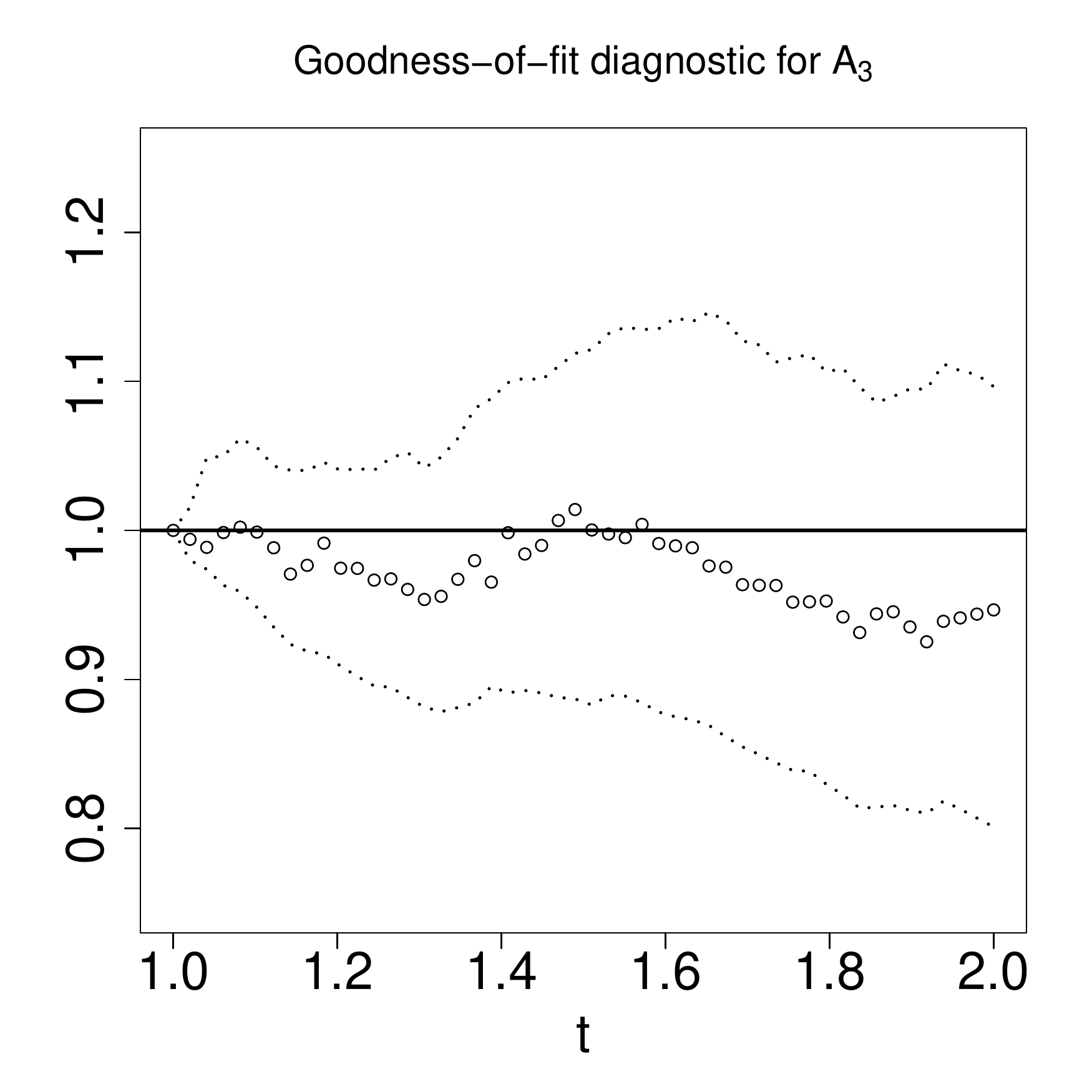}}
\caption{Abisko precipitation data: Ratio \eqref{eq:ratio}
with $u = 24$. Approximate $95 \%$ pointwise confidence intervals are obtained by bootstrapping from $\{\bm{Y}_i: i=1,\ldots, \bm{Y}_n\}$.}
\label{fig:gof}
\end{figure}

For the dependence structure, using Equation~\eqref{eq:stability} (see also Section~\ref{sec:diagnostics}) and $\gamma = 0$, we display the empirical counterpart of the ratio
\begin{equation}\label{eq:ratio}
\frac{\mathbb{P} [ \bm{Y} - \bm{u} \in A \mid \bm{y} \nleq \bm{u}]}{t \, \mathbb{P} [ \bm{Y} - \bm{u} - \bm{\sigma} \log t \in A \mid \bm{Y} \nleq \bm{u}]},
\end{equation}
where $\bm{\sigma}$ is the vector of scale parameter estimates of the marginal GP models above $u = 24$ for the sets $A_j = \{\bm{x} \in \mathbb{R}^3 : x_j > 0\}, j \in \{1,2,3\}$ (Figure~\ref{fig:gof}). The plots indicate that a GP dependence structure is appropriate. The plot for $A_1$ uses few observations and hence is more variable.

Formulas for pairwise and trivariate $\chi$ and comparisons with their empirical counterpart can be found in Section~F of the supplementary material. The model-based estimates of exceedance probabilities are $\mathbb{P}[X_1 > 0] = 0.34$ $(0.03)$, $\mathbb{P}[X_2 > 0] = 0.63$ $(0.03)$ using  values from the top row in Table~\ref{tab:fit} and delta method standard errors. The empirical probabilities are $0.32$ and $0.69$ respectively. Plots of the empirical probabilities for a range of different thresholds (not shown) confirm the chosen threshold value $u = 24$.

The test statistic in \citet[Corollary~2.5]{einmahl2016nr2} compares the estimates of $(\chi_{12},\chi_{13},\chi_{23},\chi_{123})$ with an empirical estimator. It depends on a value $k$ which represents a threshold: a low value of $k$ corresponds to a high threshold. Asymptotically the test statistic has a chi-square distribution with $2$ degrees of freedom whose $95 \%$ quantile is   $5.99$. For $k \in \{50,75,100,125,150\}$ we obtain the values $1.08$, $4.48$, $1.17$, $5.42$, and $0.99$,
and hence cannot reject the structured components model for any value of $k$.

\section{Parametric models}
\label{sec:examples}

Here we derive the explicit densities for a number of GP models. To control bias when fitting a multivariate GP distribution to threshold excesses, we often need to use censored likelihood (Section~\ref{sec:inference}) and thus not just to be able to calculate densities, but also integrals of those densities. Whilst any (continuous) distribution may be used as generator, this requirement together with the considerations in the beginning of Section~\ref{sec:models} guide our choice of models presented below. For each model we give the uncensored densities in the  subsequent subsections, and their censored versions are given in the supplementary material. The supplementary material also contains calculations of the bivariate tail dependence coefficients $\chi_{1:2}$, where these are available in closed form.



In Sections~\ref{sec:indep} and~\ref{sec:mvn} we consider particular instances of densities $f_{\bm{T}}$ and $f_{\bm{U}}$ to evaluate the corresponding densities $h_{\bm{T}}$ and $h_{\bm{U}}$ in \eqref{eq:casef} and \eqref{eq:caseg}. As noted in Section~\ref{sec:models}, even if $f_{\bm{T}} = f_{\bm{U}}$, the GP densities $h_{\bm{T}}$ and $h_{\bm{U}}$ are still different in general. Thus we will focus on the density of a random vector $\bm{V}$, denoted $f_{\bV}$, and create two GP models per $f_{\bm{V}}$ by setting $f_{\bm{T}}=f_{\bm{V}}$ and then $f_{\bm{U}}=f_{\bm{V}}$, in the latter case with the restriction $\E[e^{U_j}] < \infty$. The support for each GP density given in Sections~\ref{sec:indep} and~\ref{sec:mvn} is $\{\bx\in\mathbb{R}^d: \bx\not\leq \bzero \}$, and for brevity, we omit the indicator $\I\{ \max(\bx) > 0 \}$. In Section~\ref{sec:structured} we exhibit a construction of $h_{\bm{R}}$ in \eqref{eq:realdens}, with support depending on $\bm{\gamma}$ and $\bm{\sigma}$. In the supplementary material, we show scatterplots for some of these models together with the corresponding density contours.

In all models, identifiability issues occur if $\bm{T}$ or $\bm{U}$ have unconstrained location parameters $\bm{\beta}$, or if $\bm{R}$ has unconstrained scale parameters $\bm{\lambda}$. Indeed, replacing $\bm{\beta}$ or $\bm{\lambda}$ by $\bm{\beta}+k$ or $c\bm{\lambda}$, respectively, with $k \in \mathbb{R}$ and $c>0$, leads to the same GP distribution \citep[Proposition 1]{rootzen2016}. A single constraint, such as fixing the first parameter in the parameter vector, is sufficient to restore identifiability.

\subsection{Generators with independent components}
\label{sec:indep}

Let $\bm{V} \in \mathbb{R}^d$ be a random vector with independent components and density $f_{\bm{V}} (\bm{v}) = \prod_{j=1}^d f_{j}(v_j)$, where $f_j$ are densities of real-valued random variables. The dependence structure of the associated GP distributions is determined by the relative heaviness of the tails of the $f_j$: roughly speaking, if components have high probability of taking very different values, then  dependence is weaker than if all components have a high probability of taking similar values. Throughout, $\bm{x} \in \mathbb{R}^d$ is such that $\max( \bm{x} ) > 0$.

\noindent
{\bf Generators with independent Gumbel components:}
Let
\begin{equation*}
  f_j(v_j)
  =
  \alpha_j \exp\{-\alpha_j(v_j -\beta_j)\}
  \exp[-\exp\{-\alpha_j(v_j -\beta_j)\}], \qquad \alpha_j>0, \, \beta_j\in\mathbb{R}.
\end{equation*}
{\bf Case $f_{\bm{T}} = f_{\bm{V}}$.}
Density~\eqref{eq:casef} is
\begin{align}
  h_{\bm{T}} (\bx;\bm{1},\bm{0}) = e^{- \max(\bm{x})} \int_0^\infty  t^{-1} \prod_{j=1}^d \alpha_j \left(t e^{x_j-\beta_j}\right)^{-\alpha_j} e^{-(t e^{x_j-\beta_j})^{-\alpha_j}} \diff t. \label{eq:modconstrgumb}
\end{align}
If $\alpha_1=\ldots=\alpha_d = \alpha$ then the integral can be explicitly evaluated:
\[
  h_{\bm{T}} (\bx;\bm{1},\bm{0}) =  e^{- \max(\bm{x}) }\alpha^{d-1}\frac{\Gamma(d) \prod_{j=1}^d e^{-\alpha (x_j-\beta_j)} }{\left(\sum_{j=1}^d {e^{-\alpha (x_j-\beta_j)}}\right)^{d}}.
\]
{\bf Case $f_{\bm{U}} = f_{\bm{V}}$.}
The marginal expectation of the exponentiated variable is
$  \E[e^{U_j}] =  e^{\beta_j} \Gamma(1-1/\alpha_j)$ for $\alpha_j>1$ and $\E[e^{U_j}] = \infty$ for $\alpha_j \leq 1$. For $\min_{1\leq j \leq d} \alpha_j >1$, density~\eqref{eq:caseg} is
\begin{align}
  h_{\bm{U}} (\bx;\bm{1},\bm{0})
  =
  \frac%
    {\int_0^\infty
 \prod_{j=1}^d
	\alpha_j \,
	\left( t e^{x_j-\beta_j} \right)^{-\alpha_j}
	e^{-(t e^{x_j-\beta_j})^{-\alpha_j}}
     \diff t}%
    {\int_0^\infty \left( 1-\prod_{j=1}^d e^{-(t/e^{\beta_j})^{-\alpha_j}} \right) \diff t}.
  \label{eq:modgumb}
\end{align}

\noindent
If $\alpha_1=\ldots=\alpha_d = \alpha$ then this simplifies to:
\[
 h_{\bm{U}}(\bx;\bm{1},\bm{0}) = \frac{\alpha^{d-1} \Gamma(d-1/\alpha) \prod_{j=1}^d e^{-\alpha(x_j-\beta_j)}}{\left(\sum_{j=1}^d {e^{-\alpha(x_j-\beta_j)}}\right)^{d-1/\alpha} \Gamma(1-1/\alpha)\left(\sum_{j=1}^d e^{\beta_j\alpha}\right)^{1/\alpha}}.
\]
Observe that if in addition to $\alpha_1=\ldots=\alpha_d = \alpha$, also $\beta_1=\ldots=\beta_d = 0$, then this is the multivariate GP distribution associated to the well-known \emph{logistic} max-stable distribution.

\noindent
{\bf Generators with independent reverse Gumbel components:}
Let
\begin{equation*}
f_j(v_j) = \alpha_j\exp\{\alpha_j(v_j -\beta_j)\}
\exp[-\exp\{\alpha_j(v_j -\beta_j)\}], \qquad \alpha_j>0,\,\beta_j\in\mathbb{R}.
\end{equation*}
As the Gumbel case leads to the multivariate GP distribution associated to the logistic max-stable distribution, when $f_{\bm{U}} = f_{\bm{V}}$, the reverse Gumbel leads to the multivariate GP distribution associated to the \emph{negative logistic} max-stable distribution\footnote{The authors are grateful to Cl\'ement Dombry for having pointed out this connection.}. Calculations are very similar to the Gumbel case, and hence omitted.

\noindent
{\bf Generators with independent reverse exponential components:} Let
\begin{equation*}
  f_j(v_j) = \alpha_j\exp\{\alpha_j(v_j +\beta_j)\}, \qquad
  v_j \in (-\infty,-\beta_j), \, \alpha_j>0, \, \beta_j\in\mathbb{R}.
\end{equation*}
{\bf Case $f_{\bm{T}} = f_{\bm{V}}$.}
Density~\eqref{eq:casef} is
\begin{align}
 h_{\bm{T}} (\bx;\bm{1},\bm{0})  & =  e^{- \max(\bm{x})}\int_0^{e^{-\max(\bm{x}+\bm{\beta})}} t^{-1} \prod_{j=1}^d \alpha_j(te^{x_j+\beta_j})^{\alpha_j}   \diff t \notag\\
 &= \frac{e^{- \max(\bm{x})-\max(\bm{x}+\bm{\beta})\sum_{j=1}^d \alpha_j}}{\sum_{j=1}^d \alpha_j} \prod_{j=1}^d \alpha_j (e^{x_j+\beta_j})^{\alpha_j}. \label{eq:modconstrrevexp}
\end{align}
{\bf Case $f_{\bm{U}} = f_{\bm{V}}$.}
The expectation of the exponentiated variable is $\E[e^{U_j}] = 1/ \left\{e^{\beta_j}(1/\alpha_j+1)\right\}$, which is finite for all permitted parameter values. Density~\eqref{eq:caseg} is
\begin{align}
 h_{\bm{U}} (\bx;\bm{1},\bm{0}) &= \frac{1}{\E[e^{\max(\bm{U})}]}\int_0^{e^{-\max(\bm{x}+\bm{\beta})}}  \prod_{j=1}^d \alpha_j(te^{x_j+\beta_j})^{\alpha_j}  \diff t \notag\\
 &= \frac{(e^{-\max(\bm{x}+\bm{\beta})})^{\sum_{j=1}^d \alpha_j + 1}}{\E[e^{\max(\bm{U})}]}\frac{1}{1+\sum_{j=1}^d \alpha_j} \prod_{j=1}^d \alpha_j (e^{x_j+\beta_j})^{\alpha_j}. \label{eq:modrevexp}
\end{align}

The normalization constant may be evaluated as
\begin{align*}
\E[e^{\max(\bm{U})}]
& = \int_0^\infty \left(1-{\textstyle\prod_{j=1}^d} \min(e^{\beta_j} t, 1)^{\alpha_j}\right) \diff t \\
 & = e^{-\beta_{(d)}}- \frac{\prod_{j=1}^d e^{\alpha_j\beta_j}}{\sum_{j=1}^d \alpha_j + 1} e^{-\beta_{(1)}(\sum_{j=1}^d \alpha_j+1)} \\
 & \qquad + \sum_{i=1}^{d-1} \frac{\prod_{j=i+1}^d e^{\alpha_{[j]}\beta_{(j)}}}{\sum_{j=i+1}^d \alpha_{[j]} + 1} \left(e^{-\beta_{(i+1)}(\sum_{j=i+1}^d \alpha_{[j]} +1)}- e^{-\beta_{(i)}(\sum_{j=i+1}^d \alpha_{[j]} +1)}\right),
\end{align*}
where $\beta_{(1)}>\beta_{(2)}> \cdots > \beta_{(d)}$ and where $\alpha_{[j]}$ is the component of $\bm{\alpha}$ with the same index as $\beta_{(j)}$ (thus the $\alpha_{[j]}$s are not ordered in general). As far as we are aware, the associated max-stable model is not well known. If $\bm{\beta} = \beta \bm{1}$, then $\E[e^{\max(\bm{U})}] = [e^{-\beta}\sum_{j=1}^d \alpha_j]/[1+\sum_{j=1}^d \alpha_j]$, and $h_{\bm{U}} = h_{\bm{T}}$.

\noindent
{\bf Generators with independent log-gamma components:}
if  $e^{V_j} \sim$ Gamma$(\alpha_j,1)$ then
\begin{equation*}
  f_j (v_j) = \exp (\alpha_j v_j) \exp \{ - \exp (v_j) \} / \Gamma(\alpha_j),
  \qquad \alpha_j > 0, \, v_j \in (-\infty,\infty).
\end{equation*}
{\bf Case $f_{\bm{T}} = f_{\bm{V}}$.}
Density~\eqref{eq:casef} is
\begin{align*}
 h_{\bm{T}} (\bx;\bm{1},\bm{0})
 & = e^{-\max(\bm{x})} \prod_{j=1}^d \left( \frac{e^{\alpha_j x_j}}{\Gamma (\alpha_j)} \right) \int_0^ \infty t^{\sum_{j=1}^d \alpha_j - 1} e^{- t \sum_{j=1}^d e^{x_j}} \diff t \\
  & =  \frac{\Gamma\left(\sum_{j=1}^d \alpha_j\right)}{\prod_{j=1}^d \Gamma (\alpha_j)} \frac{ e^{\sum_{j=1}^d \alpha_j x_j - \max(\bm{x})}}{(\sum_{j=1}^d e^{x_j})^{\sum_{j=1}^d \alpha_j}}.
\end{align*}
{\bf Case $f_{\bm{U}} = f_{\bm{V}}$.}
The marginal expectation of the exponentiated variable is $\E[e^{U_j}] = \alpha_j$, hence finite for all permitted parameter values. Density~\eqref{eq:caseg} is
\begin{align*}
  h_{\bm{U}} (\bx;\bm{1},\bm{0})
  &=
  \frac{1}{\E[e^{\max(\bm{U})}]}
  \prod_{j=1}^d
    \left( \frac{e^{\alpha_j x_j}}{\Gamma(\alpha_j)} \right)
    \int_0^\infty t^{\sum_{j=1}^d \alpha_j} e^{- t \sum_{j=1}^d e^{x_j}} \diff t \\
  &=
  \frac{1}{\E[e^{\max(\bm{U})}]}
  \frac{\Gamma\left(\sum_{j=1}^d \alpha_j+1\right)}{\prod_{j=1}^d \Gamma (\alpha_j)}
  \frac{ e^{\sum_{j=1}^d \alpha_j x_j - \max(\bm{x})}}{(\sum_{j=1}^d e^{x_j})^{\sum_{j=1}^d \alpha_j+1}}.
\end{align*}
The normalization constant is
\begin{align*}
 \E[e^{\max(\bm{U})}]& =  \frac{\Gamma \left( \sum_{j=1}^d \alpha_j + 1\ \right)}{\prod_{j=1}^d \Gamma (\alpha_j)} \int_{\Delta_{d-1}} \max (u_1,\ldots,u_d) \prod_{j=1}^d u_j^{\alpha_j - 1} \diff u_1 \cdots \diff u_{d-1},
\end{align*}
where $ \Delta_{d-1} = \{(u_1,\ldots,u_d) \in [0,1]^d : u_1 + \cdots + u_d = 1\} $ is the unit simplex, and the integral can be easily computed using the \textsf{R} package \textsf{SimplicialCubature}.
This GP distribution is associated to the  Dirichlet max-stable distribution \citep{coles1991, segers2012}.

\subsection{Generators with multivariate Gaussian components}
\label{sec:mvn}

Let $f_{\bm{V}}(\bm{v}) = (2\pi)^{-d/2}|\Sigma|^{-1/2}\exp\{-(\bm{v}-\bm{\beta})^T\Sigma^{-1}(\bm{v}-\bm{\beta})/2\}$, where $\bm{\beta}\in\mathbb{R}^d$ is the mean parameter and $\Sigma\in\mathbb{R}^{d\times d}$ is a positive-definite covariance matrix. As before, $\max( \bm{x} ) > 0$. For calculations, it is simplest to make the change of variables $s=\log t$ in \eqref{eq:casef} and~\eqref{eq:caseg}.

\noindent
{\bf Case $f_{\bm{T}} = f_{\bm{V}}$.}
Density~\eqref{eq:casef} is
\begin{align}
 h_{\bm{T}} (\bx;\bm{1},\bm{0}) &=  e^{-\max(\bm{x})} \int_{-\infty}^\infty \frac{(2\pi)^{-d/2}}{|\Sigma|^{1/2}}\exp\left\{-\tfrac{1}{2}(\bx-\bm{\beta}-s \bm{1})^T\Sigma^{-1}(\bx-\bm{\beta}-s \bm{1})\right\} \diff s \notag \\
 &= \frac{(2\pi)^{(1-d)/2}|\Sigma|^{-1/2}}{(\bm{1}^T\Sigma^{-1}\bm{1})^{1/2}} \exp\left\{-\tfrac{1}{2}(\bx-\bm{\beta})^T A (\bx-\bm{\beta})- \max(\bm{x})\right\} \label{eq:modconstrloggauss}
\end{align}
with
\begin{equation}
\label{eq:A}
 A=\Sigma^{-1} - \frac{\Sigma^{-1}\bm{1}\bm{1}^T\Sigma^{-1}}{\bm{1}^T\Sigma^{-1}\bm{1}},
\end{equation}
a $d\times d$ matrix of rank $d-1$.

\noindent
{\bf Case $f_{\bm{U}} = f_{\bm{V}}$.}
The expectation $\E[ e^U_j ] = e^{\beta_j + \Sigma_{jj}/2}$ is finite for all permitted parameter values, where $\Sigma_{jj}$ denotes the $j$th diagonal element of $\Sigma$. Density~\eqref{eq:caseg} is
\begin{align*}
 h_{\bm{U}} (\bx;\bm{1},\bm{0}) &=  \frac{1}{\E[e^{\max(\bm{U})}]} \int_{-\infty}^\infty \frac{(2\pi)^{-d/2}}{|\Sigma|^{1/2}}\exp\left\{-\tfrac{1}{2}(\bx-\bm{\beta}-s \bm{1})^T\Sigma^{-1}(\bx-\bm{\beta}-s \bm{1}) - s\right\} \diff s\\
 &= \frac{(2\pi)^{(1-d)/2}|\Sigma|^{-1/2}}{\E[e^{\max(\bm{U})}](\bm{1}^T\Sigma^{-1}\bm{1})^{1/2}} \exp\left\{-\tfrac{1}{2}\left[(\bx-\bm{\beta})^T A (\bx-\bm{\beta}) + \frac{2(\bx-\bm{\beta})^T\Sigma^{-1}\bm{1}-1}{\bm{1}^T\Sigma^{-1}\bm{1}}\right] \right\},
\end{align*}
with $A$ as in \eqref{eq:A}. This is the GP distribution associated to the Brown--Resnick or H\"{u}sler--Reiss max-stable model \citep{kabluchko2009,husler1989}. A variant of the density formula with $\E[e^{U_j}] = 1$ (equivalently $\bm{\beta} = -\mbox{diag}(\Sigma)/2$) was given in \citet{wadsworth2014}. The normalization constant is $\int_0^\infty \left[1-\Phi_d(\log t \bm{1}-\bm{\beta};\Sigma)\right] \diff t$, where $\Phi_d(\cdot;\Sigma)$ is the zero-mean multivariate normal distribution function with covariance matrix $\Sigma$. This normalization constant can be expressed as a sum of multivariate normal distribution functions \citep{huser2013}.

\subsection{Generators with structured components}
\label{sec:structured}

We present a model for $\bm{R}$ based on cumulative sums of exponential random variables and whose components are ordered; for the components of the corresponding GP vector to be ordered as well, we assume that $\bm{\gamma} = \gamma \bm{1}$ and $\bm{\sigma} = \sigma \bm{1}$. We restrict our attention to $\gamma \in [0,\infty)$ in view of the application we have in mind: this model is used in Section~\ref{sec:rainfall} to model cumulative precipitation amounts which may trigger landslides.

\noindent
{\bf Case $\bm{\gamma} = \bm{0}$.}
By construction, the densities $h_{\bm{R}}(\,\cdot \, ; \bm{1},\bm{0})$ and $h_{\bm{U}} (\,\cdot \, ; \bm{1},\bm{0})$ coincide since $\bm{R} = \bm{U}$. Let $\bm{R} \in (-\infty,\infty)^d$ be the random vector whose components are defined by
\begin{equation*}
R_j = \log \left( {\textstyle\sum}_{i=1}^j E_i \right), \qquad E_j \overset{\mathrm{iid}}{\sim} \mathrm{Exp} (\lambda_j), \qquad j= 1,\ldots,d,
\end{equation*}
where the $\lambda_j$ are the mean values of the exponential distributions. Its density, $f_{\bm{R}}$, is
\begin{equation*}
f_{\bm{R}}(\bm{r}) = \begin{cases}
\left( \prod_{j=1}^d  \lambda_j e^{r_j} \right) \exp \left\{ - \sum_{j=1}^{d} (\lambda_j - \lambda_{j+1} ) e^{r_j} \right\}, & \text{ if } r_1 < \ldots < r_d, \\
0, & \text{ otherwise},
\end{cases}
\end{equation*}
where we set $\lambda_{d+1} = 0$. In view of \eqref{eq:caseg}, $R_1 < \ldots < R_d$ (or equivalently $U_1 < \ldots < U_d$) implies $X_{0,1} < \ldots  < X_{0,d}$. The density of $\bX_0$ is given as follows: if $x_1 < \ldots < x_d$, then
\begin{align}\label{eq:orderdens}
h_{\bm{R}} ( \bx ; \bm{1},\bm{0}) & = \frac{\mathbbm{1} \left( x_d > 0 \right)}{\E [ e^{R_d}]} \left( \prod_{j=1}^d \lambda_j e^{x_j} \right) \int_0^\infty t^d \exp \left\{ - t \left( \sum_{j=1}^d (\lambda_j - \lambda_{j+1} ) e^{x_j} \right) \right\} \diff t \notag \\
& =  \frac{\I(x_d > 0) \, d! \, \prod_{j=1}^d \lambda_j e^{x_j} }{\left( \sum_{j=1}^d \lambda_j^{-1} \right) \left( \sum_{j=1}^d (\lambda_j - \lambda_{j+1} ) e^{x_j} \right)^{d+1}},
\end{align}
while $h_{\bm{R}}( \bx ; \bm{1}, \bm{0})$ is zero otherwise. The density $h_{\bm{R}} (\bx ; \bm{\sigma},\bm{0})$ is  obtained from~\eqref{eq:mgpdens}.

\noindent
{\bf Case $\bm{\gamma} > \bm{0}$.}
Let $\bm{R} \in (0,\infty)^d$ be the random vector whose components are defined by
\begin{equation*}
R_j = \sum_{i=1}^j E_i, \qquad E_j \overset{\mathrm{iid}}{\sim} \mathrm{Exp} (\lambda_j), \qquad j= 1,\ldots,d,
\end{equation*}
 Its density, $f_{\bm{R}}$, is similar to the one for $\bm{\gamma} = \bm{0}$. Then
\begin{equation*}
\E \left[ e^{\max (\bm{U})} \right] = \E \left[ \max_{1 \leq j \leq d} \left( \frac{\gamma R_j}{\sigma} \right)^{1/\gamma} \right]  = \left( \frac{\gamma}{\sigma} \right)^{1/\gamma} \E \left[ R_d^{1/\gamma} \right].
\end{equation*}
The distribution of $R_d$ is called generalized Erlang if $\lambda_i \neq \lambda_j$ for all $i \neq j$ \citep{neuts1975}, and, letting $f_{R_d}$ denote its density we get
\begin{align*}
 \E \left[ R_d^{1/\gamma} \right]  =  \int_0^{\infty} r^{1/\gamma} f_{R_d} (r) \diff r
 = \Gamma \left(\frac{1}{\gamma} + 1 \right) \sum_{i=1}^d \lambda_i^{-1/\gamma} \left(\prod_{j=1, j \neq i}^d \frac{\lambda_j}{\lambda_j-\lambda_i} \right).
\end{align*}
If $\lambda_1 = \ldots = \lambda_d$, then $R_d$ follows an Erlang distribution. By \eqref{eq:realdens}, the density of $\bX$ becomes, for $x_d > \ldots > x_1 > - \sigma/\gamma$ and $x_d > 0$,
\begin{align*}
h_{\bm{R}} (\bx ; \bm{\sigma}, \bm{\gamma}) & = \frac{\left(\prod_{j=1}^d \lambda_j\right) \int_0^\infty t^{d \gamma} \exp \left\{ - t^{\gamma} \sum_{j=1}^d (\lambda_j - \lambda_{j+1} ) (x_j + \sigma/\gamma) \right\} \diff t}{\left( \frac{\gamma}{\sigma} \right)^{1/\gamma} \E \left[ R_d^{1/\gamma} \right]} \\
& = \frac{\left( \prod_{j=1}^d \lambda_j \right)  \left( \frac{\gamma}{\sigma} \right)^{-1/\gamma} \Gamma \left( d + \frac{1}{\gamma} \right) / \Gamma \left(\frac{1}{\gamma} \right)}{ \left( \sum_{j=1}^d (\lambda_j - \lambda_{j+1}) x_j + (\sigma/\gamma) \lambda_1 \right)^{d + 1/\gamma}\sum_{i=1}^d \lambda_i^{-1/\gamma} \left(\prod_{j=1, j \neq i}^d \frac{\lambda_j}{\lambda_j-\lambda_i} \right)}.
\end{align*}

\section*{Acknowledgements}

The authors gratefully acknowledge support from: the Knut and Alice Wallenberg foundation (Kiriliouk, Rootz\'en, Wadsworth); ``Projet d'Act\-ions de Re\-cher\-che Concert\'ees'' No.\ 12/17-045 of the ``Communaut\'e fran\c{c}aise de Belgique'' (Kiriliouk, Segers); IAP research network grant P7/06 of the Belgian government (Segers); EPSRC fellowship grant EP/P002838/1 (Wadsworth). Finally, we thank the Abisko Scientific Research Station for access to their rainfall data.

\medskip
\begin{center}
{\large\bf SUPPLEMENTARY MATERIAL}
\end{center}

\begin{description}

\item[Supporting information:] Details of censored likelihoods, simulation study, and additional information relating to the analyses in Sections~\ref{sec:banks} and~\ref{sec:rainfall}. (.pdf)
\item[Code and data:] Code and data for the analyses in Sections~\ref{sec:banks} and~\ref{sec:rainfall}, with description (.zip)



\end{description}

\bibliographystyle{apalike}
\bibliography{libraryMGPD}

\end{document}